# Effect of a magnetic field up to 9 T on the temperature dependence of the pseudogap in YBa$_2$Cu$_3$O$_{7-\delta}$ films

A. S. Kolisnyk[1], M. V. Shytov[1], E. V. Petrenko[1], A. V. Terekhov[1], L. V. Bludova[1],
A. Sedda[2], E. Lähderanta[2], and A. L. Solovjov[1,2,3]

[1]*B. Verkin Institute for Low Temperature Physics and Engineering of the National Academy of Sciences of Ukraine Kharkiv 61103, Ukraine*

[2]*LUT University, Department of Physics, Lappeenranta 53850, Finland*

[3]*Institute for Low Temperatures and Structure Research of PAS, Wroclaw 50-422, Poland*

E-mail: solovjov@ilt.kharkov.ua

The work analyzes the effect of a magnetic field $B$ directed along the $c$ axis ($B \parallel c$) up to 9 T on the resistivity $\rho(T)$, fluctuation conductivity (FLC) $\sigma'(T)$ and pseudogap $\Delta^*(T)$ in thin films of YBa$_2$Cu$_3$O$_{7-\delta}$ with a critical temperature of the superconducting transition $T_c = 88.8$ K. In contrast to previous work (*Low Temp. Phys.* **51**, 1061 (2025) [*Fiz. Nyzk. Temp.* **51**, 1180 (2025)]), where the magnetic field was directed along the $ab$ plane ($B \parallel ab$), the influence of the field on the sample is stronger due to the contribution of both spin-orbit and Zeeman effects. As expected, the magnetic field does not affect $\rho(T)$ in the normal state. However, at the superconducting (SC) transition, it sharply increases $\rho(T)$, the width of the SC transition $\Delta T_c$, and the coherence length along the $c$ axis, $\xi_c(0)$, but at the same time reduces both $T_c$ and the range of the SC fluctuations $\Delta T_{fl}$. The FLC reveals a transition at a characteristic temperature $T_0$ from the three-dimensional Aslamazov–Larkin (3D–AL) theory near $T_c$ to the two-dimensional Maki–Thompson (2D–MT) fluctuation theory. However, at $B = 1$ T, the 2D–MT contribution is completely suppressed, and above $T_0$, $\sigma'(T)$ is unexpectedly described by the 2D–AL fluctuation contribution, indicating the formation of a two-dimensional vortex lattice in the film under the action of a magnetic field. It was found that the BEC–BCS transition temperature, $T_{pair}$, which corresponds to the maximum of the $\Delta^*(T)$ dependence, shifts to the region of lower temperatures with increasing $B$, and the maximum value of $\Delta^*(T_{pair})$ decreases in fields $B > 5$ T. It was found that with increasing field, the low-temperature maximum near $T_0$ is smeared and disappears at $B > 1$ T. In addition, above the Ginzburg temperature $T_G$, for $B > 1$ T, a minimum appears on $\Delta^*(T)$ at $T_{min}$, which becomes very pronounced with a subsequent increase in $B$. As a result, the overall value of $\Delta^*(T_G)$ decreases noticeably, most likely due to the pair-breaking effect. At the same time, $\Delta T_{fl}$ and $\xi_c(0)$ increase sharply by approximately 3 times with increasing $B$ above 1 T. Our results confirm the possibility of the formation of a vortex state in YBa$_2$Cu$_3$O$_{7-\delta}$ by a magnetic field and its evolution with increasing $B$.

Keywords: high-temperature superconductors, YBCO films, excess conductivity, fluctuation conductivity, magnetic field, coherence length, pseudogap.

## 1. Introduction

One of the key tasks of modern solid-state physics remains the development of a theory capable of comprehensively describing all the features of high-temperature superconductors (HTSCs). Unfortunately, a serious obstacle is still the lack of a clear understanding of the physics of internal interactions in multicomponent compounds such as HTSCs, in particular, the mechanism of superconducting (SC) pairing, which provides the possibility of achieving a superconducting transition temperature $T_c > 100$ K. For more than three decades since the discovery of HTSCs, progress in this direction has remained limited. This is due to the high complexity of the crystal and electronic structure of HTSCs, the variety of phase states, as well as the strong dependence of properties on the level of doping, structural defects, and external influences. In recent years, special attention has been paid to the study of the normal state of HTSCs, in which unique phenomena appear, in particular the pseudogap (PG) state [1–7], the nature of which has not yet been fully elucidated.

Among the representatives of HTSCs, cuprates occupy a special place, i.e., copper oxides with active planes $CuO_2$, which demonstrate a unique combination of properties: high $T_c$ values, the presence of PG, low concentration of charge carriers $n_f$, strong electronic correlations, and pronounced anisotropy of conductivity [8–11]. For the compound $YBa_2Cu_3O_{7-\delta}$ (YBCO), it is characteristic that the coherence length in the $ab$ plane, $\xi_{ab}(T)$, is approximately ten times greater than the coherence length $\xi_c(T)$ along the $c$ axis [12]. This reflects its quasi-two-dimensional electronic structure and causes differences in behavior under the influence of a magnetic field oriented in different directions. The low concentration of charge carriers together with the quasi-two-dimensional structure of HTSCs creates conditions for the formation of local pairs (LPs) [10, 13] below the pseudogap opening temperature $T^* \gg T_c$, which are probably responsible for the formation of the PG state ([12–15] and references therein). In the temperature range $T \leq T^*$, such pairs exist as strongly bound bosons (SBBs), which obey the Bose–Einstein condensation theory [14], and have a small size ($\xi_{ab}(T^*) \approx 10$ Å) and high binding energy, which makes them resistant to thermal and magnetic interactions. As the temperature decreases, the coherence length $\xi(T)$ increases, and the SBBs gradually transform into fluctuating Cooper pairs (FCPs), which obey the BCS theory near $T_c$ [13–15]. Condensation of FCPs into a superconducting state is possible only from the three-dimensional (3D) state [9, 11], which is realized under the condition $\xi_c(T) > d$, where $d$ is the size of the YBCO unit cell along the $c$ axis. Therefore, when $\xi_c(T) > d$ near $T_c$, the quasi-two-dimensional (2D) state of high-temperature superconductor (HTSC) always transforms into a three-dimensional one [12, 16].

Nevertheless, other models have been proposed to explain the physics of the PG, such as spin fluctuations [17], charge density waves (CDW) [18, 19], and spin density waves (SDW) [4, 18, 20], charge ordering (CO) ([4, 18, 21, 22] and references therein). However, despite the fact that interest in studying the PG has increased significantly in recent years [5, 23–26], the physics of the PG state is still not fully understood, and, therefore, the cause of the PG opening has not been resolved. According to the latest ideas [4, 15, 20], below $T^*$, a restructuring of the Fermi surface is quite possible, which largely determines the unusual properties of cuprates in the PG region. At the same time, despite the extremely large number of works devoted to the study of HTSCs and, in particular, the PG state, there is a shortage of works devoted to the influence of the magnetic field on the excess conductivity in cuprates. However, it is the study of the influence of the magnetic field on the FLC and the PG properties that allows us to answer the question: which of the above-considered physical mechanisms of PG in cuprates actually takes place?

To our knowledge, the effect of the magnetic field on the FCPs in YBCO materials has been analyzed in a relatively small number of studies to date [27–30]. However, in [27], the applied field does not exceed 1.27 T, which is too small for any unambiguous conclusions. In [28], a magnetic field of $B = 12$ T was used, and in [29], $B = 9$ T was used. However, in [28], a decrease in $\xi_c(T)$ is reported along with a decrease in $T_c$ with increasing magnetic field. This is surprising, since in the traditional theory of superconductivity, $\xi \sim 1/T_c$ [31]. On the other hand, in both works, the authors did not show the evolution of the FCPs under the influence of a magnetic field. In our previous works [32, 33], we studied the effect of magnetic field on YBCO films in a magnetic field $B$ from 0 to 8 T oriented along $ab$ plane ($B \parallel ab$). Of particular interest is the effect of an external magnetic field oriented along the $c$ axis ($B \parallel c$) on the FLC and PG in YBCO. In this configuration, the magnetic field penetrates the $CuO_2$ plane perpendicularly, which significantly enhances vortex effects, reduces the effective coherence length $\xi(T)$, and also affects the FLC modes, including the suppression of the two-dimensional Maki–Thompson (2D–MT) contribution and a change in the 3D–2D crossover temperature $T_0$. The influence of the field is accompanied by the formation of a lattice of Abrikosov vortices, which interact with each other and with fluctuating pairs, changing the dynamics of the superconducting state. It should be noted that the influence of the magnetic field on the PG and FLC in the $B \parallel c$ configuration has not been studied sufficiently. The few available works in this area indicate significant differences in the behavior of the system compared to the $B \parallel ab$ configuration, in particular, a faster decrease in $T_c$, a more pronounced expansion of the SC transition, and the appearance of additional features in the temperature dependences $\sigma'(T)$ at high fields. These effects can be directly related to the dynamics of the vortices, in particular, to the transitions between different vortex phases, which makes their study particularly relevant.

In the paper, a comprehensive study of the effect of a magnetic field of up to 9 T, oriented along the $c$ axis, on the electrical resistance $\rho(T)$, fluctuation conductivity $\sigma'(T)$ and PG $\Delta^*(T, B)$ in thin YBCO films obtained by epitaxial deposition on $(LaAlO_3)_{0.3}(Sr_2TaAlO_6)_{0.7}$ substrates was carried out. This choice of the experimental configuration makes it possible to study the evolution of the superconducting fluctuation region, the parameters of the 3D–2D transition, the behavior of the coherence length $\xi_c(T, B)$ and the temperature dependences of the PG parameter $\Delta^*(T, B)$ under conditions when the magnetic field directly interacts with charge carriers in the $CuO_2$ plane. The obtained results provide an opportunity to better understand the role of the magnetic field oriented perpendicular to the active planes of $CuO_2$ in the formation of the pseudogap state, as well as to assess the contribution of vortex dynamics to the processes of suppression of the superconducting fluctuation. This, in turn, helps to approach the solution of the fundamental problem of HTSCs physics: the elucidation of the nature of the Cooper pairing mechanism under conditions of strong electron correlation and quasi-two-dimensionality of the crystal structure.

## 2. Experiment

The epitaxial films of YBCO were deposited at $T = 770$ °C in 3 mbar oxygen pressure on (LaAlO$_3$)$_{0.3}$(Sr$_2$TaAlO$_6$)$_{0.7}$ substrates. The thickness of deposited films was approximately 100 nm and was controlled by the deposition time of respective targets [34]. X-ray diffraction analysis confirmed the high quality of the samples. All films had a well-ordered crystalline structure with the *c* axis oriented perpendicular to the CuO$_2$ planes.

Subsequently, the films were formed into clear Hall-bar structures with dimensions of 2.35 × 1.24 mm by photolithography and chemical etching. To ensure electrical contacts, gold wires were attached to the contact pads using silver epoxy paste, which allowed to achieve a contact resistance of less than 1 Ω. The measurements were performed using the Quantum Design PPMS-9 automated complex, using an alternating current with an amplitude of approximately 100 μA and a frequency of 19 Hz. Measurements of the temperature dependence of the in-plane resistivity ρ$_{ab}$(T) were carried out using a four-probe scheme. The studies were carried out in a constant magnetic field of up to 9 T, which was created by the PPMS system; the magnetic field was oriented parallel to the *c* axis.

## 3. Results and discussion

### 3.1. Resistivity

Figure 1 shows the temperature dependence of the resistivity $\rho(T) = \rho_{ab}(T)$ of the YBCO film in the absence of an external magnetic field. In the temperature range $T^*$ from 219 to 300 K, a linear dependence of ρ(T) is observed with a slope $a = d\rho/dT = 2.01$ μΩ·cm/K. The slope was determined by linear approximation of the experimentally obtained curves, which confirmed the linear nature of ρ(T) with a mean square error $\Delta\rho(T) = (0.009 \pm 0.002)$ μΩ·cm.

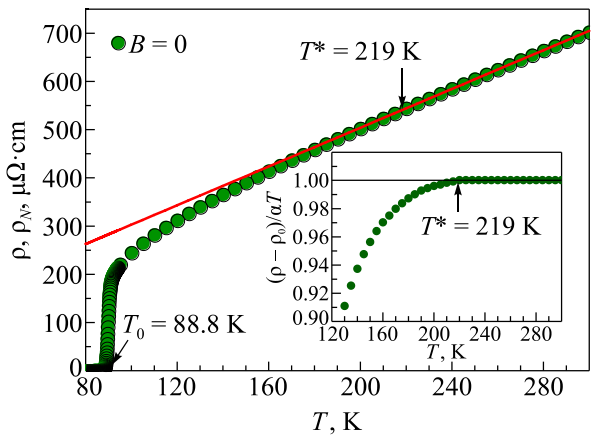

*Fig. 1.* (Color online) Temperature dependence of ρ for the YBa$_2$Cu$_3$O$_{7-\delta}$ film in the absence of an external magnetic field ($B = 0$, green dots). The solid red line determines $\rho_N(T)$ extrapolated to the low-temperature region. Inset: method for determining $T^*$ using the criterion $[\rho(T) - \rho_0]/\alpha T = 1$ [35].

The pseudogap opening temperature $T^*$, which significantly exceeds $T_c$, is defined as the point of deviation of the resistivity curve from linearity downwards (Fig. 1). A more precise approach to determining $T^*$ with an accuracy of ±1 K is to study the criterion $[\rho(T) - \rho_0]/aT = 1$ (inset in Fig. 1), where *a* is the slope of the extrapolated resistivity in the normal state $\rho_N(T)$, and $\rho_0$ is the residual resistivity determined by the point of $\rho_N(T)$ intersection with the *Y* axis [16, 35, 36].

Both methods give the same value of $T^* = 219$ K, which is typical for high-quality YBCO films with $T_c \approx 88$ K, in good agreement with literature data [16, 36].

The increase in the magnetic field $B$ leads to a significant expansion of the resistive transition (see Fig. 2), the formation of magnetoresistance $MR = [\rho(H) - \rho(0)]/\rho(0)$, and a linear decrease in $T_c$ (see Fig. 3).

It is important to note that the magnetic field does not affect the resistivity in the normal state [37, 38]. The absence of step-like features on the transitions to the superconducting state at all values of $B$ indicates the high quality of the sample, its structural homogeneity, and the absence of extraneous phases. Unfortunately, no other information can be obtained from resistive measurements. That is why the study of FLC and PG was conducted.

### 3.2. Fluctuation conductivity

As noted above, in the normal state at high temperatures $T > T^*$, the dependence $\rho_{ab}(T) = \rho(T)$ in HTSCs is linear [39]. Below $T^*$, $\rho(T)$ deviates from linearity towards lower values (Fig. 1), which leads to excess conductivity:

$$\sigma'(T) = \sigma(T) - \sigma'_N(T) = \frac{1}{\rho(T)} - \frac{1}{\rho_N(T)}$$

or $\quad \sigma'(T) = [\rho_N(T) - \rho(T)] / [\rho(T)\rho_N(T)],$ (1)

where $\rho_N(T) = aT + b$ is the resistivity of the sample in the normal state, extrapolated to the low temperature region [16, 36, 40, 41]. It should be noted that according to the local pair model [39], the linear dependence of ρ(T) above $T^*$ is the normal state of the HTSCs, which is characterized by the stability of the Fermi surface [4, 18, 20, 39].

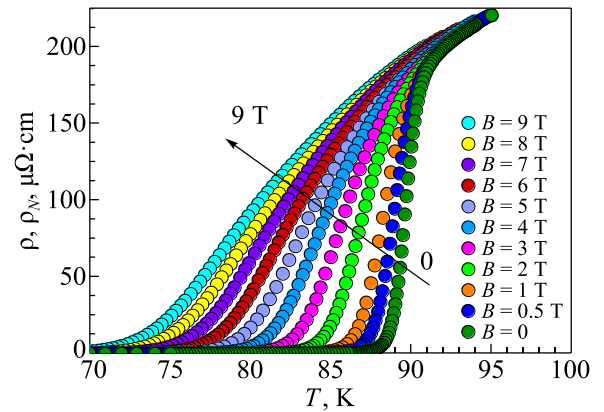

*Fig. 2.* (Color online) Temperature dependences of the resistivity of the YBa$_2$Cu$_3$O$_{7-\delta}$ film, measured in the region of the superconducting transition in a magnetic field of 0–9 T parallel to the *c* axis.

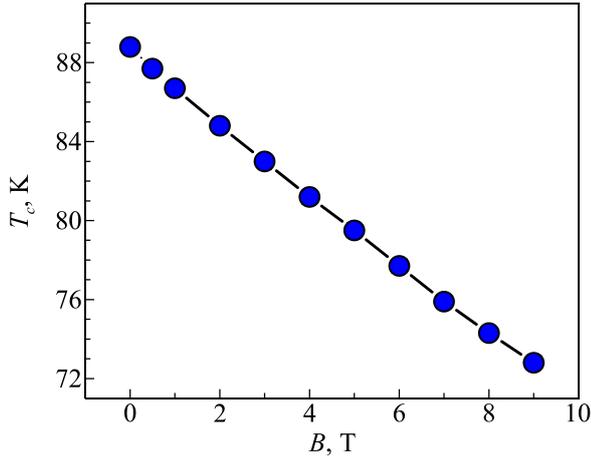

*Fig. 3.* (Color online) Dependence of the critical temperature $T_c$ of the $YBa_2Cu_3O_{7-\delta}$ film on a magnetic field of 0–9 T parallel to the $c$ axis.

According to the latest ideas, the small value of the coherence length in combination with the quasi-layer structure of the HTSCs leads to the formation of a noticeable region of superconducting fluctuations at $\rho(T)$ near $T_c$, where $\sigma'(T)$ corresponds to the classical fluctuation theories of Aslamazov–Larkin (AL) and Maki–Thompson (MT). It is important to emphasize that changes in the oxygen content, the presence of impurities, and structural defects significantly affect $\sigma'(T)$, and therefore the realization of the FLC and PG regimes above $T_c$ [16, 42–44]. The fluctuation conductivity of the studied YBCO film was determined by analyzing the excess conductivity, which was calculated by the standard method using Eq. (1). The FLC analysis was carried out within the framework of the local pair model, which assumes the presence of paired fermions in HTSCs in the temperature range $T^* > T > T_c$ [1, 13, 14, 23, 45].

The key point is the definition of the mean field temperature $T_c^{mf} > T_c$, which separates the region of superconducting fluctuations from the region of critical fluctuations near $T_c$ (where the mean field theory does not work) [31]. It is important that by defining $T_c^{mf}$, one can calculate the reduced temperature ε, which is used in all equations, according to the equation:

$$\varepsilon = \frac{T - T_c^{mf}}{T_c^{mf}}. \quad (2)$$

It is clear from this that the correct definition of $T_c^{mf}$ plays a key role in the calculations of the FLC and PG. Near $T_c$, the coherence length along the $c$ axis, $\xi_c(T)$, is larger than $d$ ($d \approx 11.7$ Å), the lattice parameter along the $c$ axis of the YBCO unit cell [46–48]. As a result, the FCPs interact throughout the volume of the sample, forming a three-dimensional (3D) state [16, 46, 48]. Thus, near $T_c$, the FLC can be described by the 3D equation of the Aslamazov–Larkin (3D–AL) theory [49, 50] with the critical exponent $\lambda = -1/2$, which defines the FLC in any 3D system:

$$\sigma'_{AL3D} = C_{3D} \frac{e^2}{32 h \xi_c(0)} \varepsilon^{-1/2}, \quad (3)$$

where $C_{3D}$ is the scaling factor. Based on the fact that $\sigma'(T) \sim \varepsilon^{-1/2}$ we obtain $\sigma'^{-2}(T) \sim \varepsilon \sim T - T_c^{mf}$, i.e., at $T = T_c^{mf}$ $\sigma'^{-2}(T)$ is equal to zero, which determines $T_c^{mf}$ (Fig. 4). Also, in Fig. 4, the arrows indicate $T_c$, the Ginzburg temperature $T_G$, down to which the fluctuation theories are valid [31, 46, 48], and the 3D–2D crossover temperature $T_0$ which limits the 3D fluctuation region. Above $T_0 = 90.9$ K (Fig. 4) at $B = 0$, the data deviate from linearity to the right, indicating the presence of a 2D Maki–Thompson (2D–MT) contribution to the FLC. If the 2D–MT contribution to the FLC [46, 48] is suppressed by defects induced by the magnetic field, the data always deviate to the left [Fig. 4(b)] [37, 51]. To detail the observed unexpected behavior, it is necessary to plot $\ln \sigma'$ as a function of $\ln \varepsilon$ for each magnetic field value and compare the results with fluctuation theories.

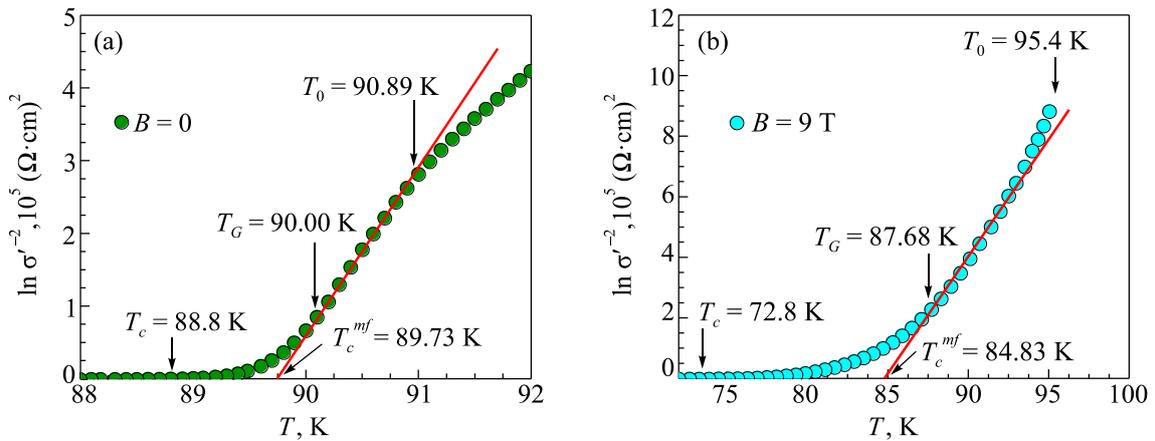

*Fig. 4.* (Color online) Dependences of $\sigma'^{-2}(T)$ for $YBa_2Cu_3O_{7-\delta}$ film (a) at $B = 0$ (green dots) and (b) $B = 9$ T (turquoise dots), which highlights the different behavior of the FLC above $T_0$ due to the influence of defects. The arrows indicate the temperatures $T_c$, $T_c^{mf}$, the Ginzburg temperature $T_G$, and the 3D–2D crossover temperature $T_0$.

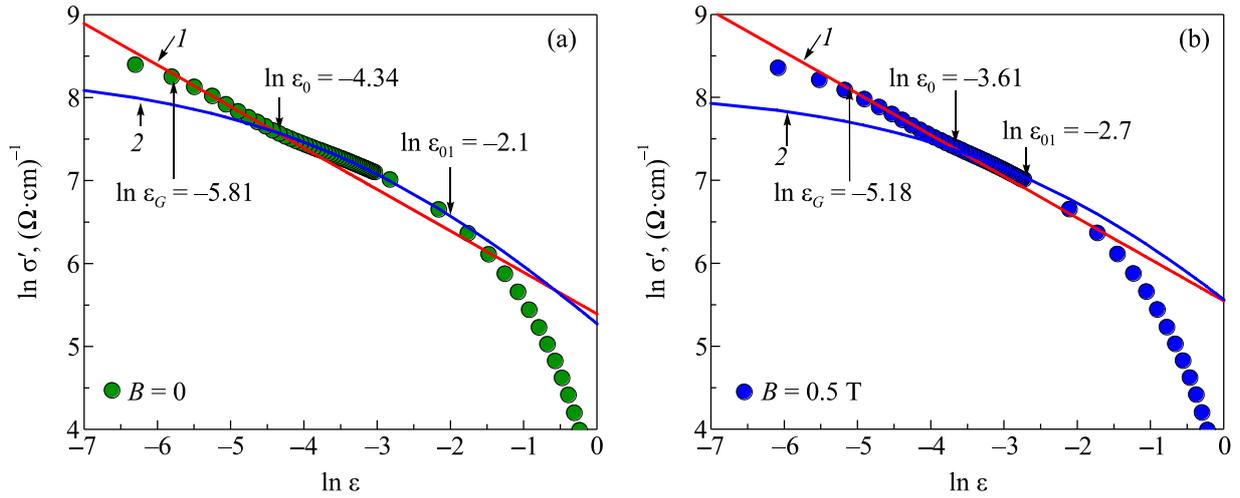

*Fig. 5.* (Color online) $\ln \sigma'$ versus $\ln \varepsilon$ for $YBa_2Cu_3O_{7-\delta}$ film (a) at $B = 0$ (green dots) and (b) 0.5 T (blue dots) compared to fluctuation theories: 3D–AL (line *1*) and 2D–MT (curve *2*); $T_0$ ($\ln \varepsilon_0$) is the 3D–2D crossover temperature, which defines the 3D region, $T_{01}$ ($\ln \varepsilon_{01}$) defines the entire region of superconducting fluctuations, and $T_G$ ($\ln \varepsilon_G$) is the Ginzburg temperature.

Having determined $T_c^{mf}$ and, consequently, the reduced temperature $\varepsilon$ [Eq. (2)], we construct the temperature dependences of the FLC in double logarithmic coordinates $\ln \sigma'(\ln \varepsilon)$ at different values of the magnetic field $B$ (Figs. 5 and 6). The corresponding dependences are shown in Fig. 5(a) for zero magnetic field and in Fig. 5(b) for a magnetic field of $B = 0.5$ T. As expected, near $T_c$, in the temperature interval $T_G - T_0$ ($\ln \varepsilon_0 = -4.34$), the FLC is well described by the fluctuation contribution of the 3D–AL [Eq. (3)]. In double logarithmic coordinates, this is a straight red line of the 3D–AL (*1*) with a slope of $\lambda = -1/2$. As mentioned above, this means that the classical three-dimensional FLC is realized in the HTSCs when $T \to T_c$ and $\xi_c(T) > d$ [41, 49, 52]. Above the crossover temperature $T_0$, $\xi_c(T) < d$ [46, 48, 49, 52], and this is no longer a 3D regime. However, as before, $\xi_c(T) > d_{01}$ ($d_{01}$ is the distance between the $CuO_2$ conducting planes). Thus, up to a temperature $T_{01}$ [$\ln \varepsilon_{01} = -2.1$, Fig. 5(a)], $\xi_c(T)$ links the internal $CuO_2$ planes through the Josephson interaction [46, 52]. This is a 2D fluctuation regime, which is approximated by the 2D–MT equation (2D–MT curve) of the MT theory modified by Hickam–Larkin for HTSCs [48]:

$$\sigma'_{2DMT} = C_{2D} \frac{e^2}{8d\hbar} \frac{1}{1-\alpha/\delta} \ln\left(\frac{\delta}{\alpha} \frac{1+\alpha+\sqrt{1+2\alpha}}{1+\delta+\sqrt{1+2\delta}}\right) \varepsilon^{-1}, \quad (4)$$

where $\alpha = 2[\xi_c(0)/d]^2 \varepsilon^{-1}$ is the coupling parameter, and

$$\delta = \beta \frac{16}{\pi h}\left[\frac{\xi_c(0)}{d}\right]^2 k_B T \tau_\phi \quad (5)$$

is the pair-breaking parameter. The factor $\beta = 1.203(\ell/\xi_{ab})$, where $\ell$ is the mean free path, corresponds to the case of the clean limit ($\ell > \xi$) being typical for HTSCs [15, 16, 53], $\tau_\varphi$ is the phase relaxation time of fluctuating pairs [48].

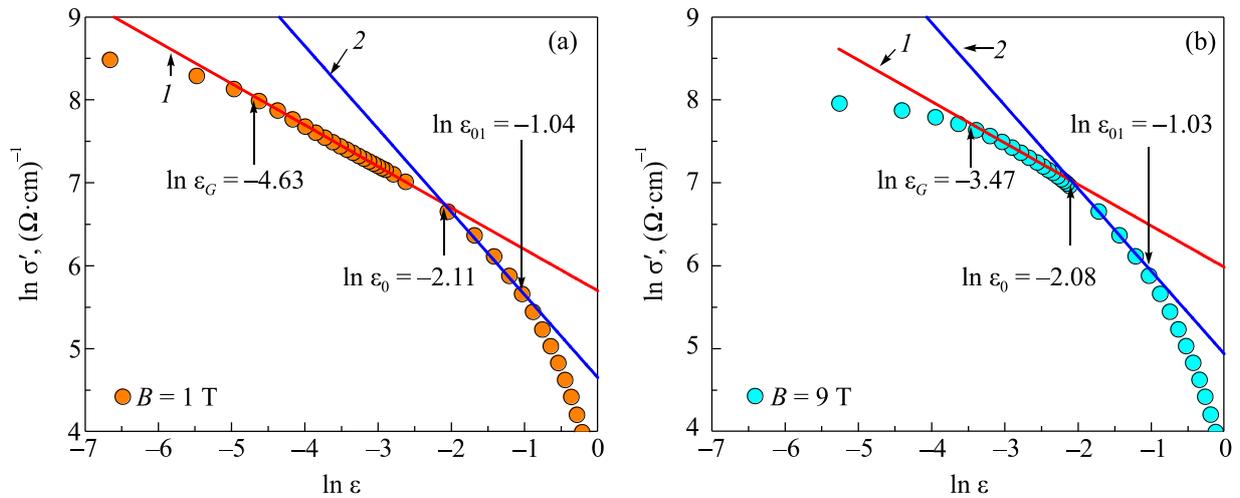

*Fig. 6.* (Color online) $\ln \sigma'$ versus $\ln \varepsilon$ for $YBa_2Cu_3O_{7-\delta}$ film (a) at $B = 1$ T (light orange dots) and (b) 9 T (turquoise dots); in both cases the 2D–MT contribution is completely suppressed by the magnetic field, and $\ln \sigma'$ above $T_0$ ($\ln \varepsilon_0$) is described by Eq. (7) 2D–AL (line *2*).

Above $T_{01}$, the experimental points deviate downward from the theory (Fig. 5), indicating that classical fluctuation theories no longer work. Thus, $T_{01}$ limits the region of SC fluctuations $\Delta T_{fl} = T_{01} - T_G$ from above, and the temperature $T_G$ limits the region of SC fluctuations from below. As a result, below $T_G$, denoted as $\ln \varepsilon_G$ in Figs. 5 and 6, the experimental points also deviate downward from the theory, indicating a transition to the region of critical fluctuations or SC order parameter $\Delta$ immediately near $T_c$, where $\Delta < kT$ [31, 49]. At $T_0$, the condition $\xi_c(T_0) = d = 11.7$ Å [47] is satisfied, which allows us to determine $\xi_c(0)$ [16, 46, 52, 53]:

$$\xi_c(0) = d\sqrt{\varepsilon_0}. \qquad (6)$$

For $\ln \varepsilon_0 = -4.34$ [Fig. 5(a)], we obtain $\xi_c(0) = 1.33$ Å ($B = 0$). This value of the coherence length is typical for well-structured HTSCs with an oxygen content close to the optimal (OD) [11]. Moreover, the obtained values of $\xi_c(0)$ and $d_{01} = 3.82$ Å are the same as those obtained in our previous work for this sample [33].

Figures 5 and 6 show the dependences of $\ln \sigma'(\ln \varepsilon)$ for magnetic fields $B = 0$, 0.5 (Fig. 5), and 1 and 9 T (Fig. 6). With increasing magnetic field, the fluctuation contribution of 2D–MT is gradually suppressed ($B = 0.5$ T). Finally, with increasing magnetic field $B > 1$ T above $T_0$ $\ln \sigma'(\ln \varepsilon)$ is now unexpectedly well described by the fluctuation contribution of 2D–AL theory [50]:

$$\sigma'_{2DAL} = C_{2D} \frac{e^2}{16\hbar d} \varepsilon^{-1}, \qquad (7)$$

where $d$ is the thickness of the sample. As can be seen from Fig. 6, this leads to a sharp increase in $T_0$ ($\ln \varepsilon_0$ in the figures), i.e., the region of 3D fluctuations, and hence $\xi_c(0)$. Indeed, the coherence length $\xi_c(0)$ increases from 1.33 Å ($B = 0$) to 4.14 Å ($B = 9$ T) (Table 1). The dependence of $\xi_c(0)$ on magnetic field is shown in Fig. 7. It is seen that $\xi_c(0)$ increases sharply by about 3 times at $B > 1$ T, and then remains almost independent of $B$. At the same time, both $T_{01}$ ($\ln \varepsilon_{01}$ in the figures), i.e., the region of the SC fluctuations $\Delta T_{fl}$, and the distance between the conductive planes of $CuO_2$, $d_{01}$, also increase significantly (Table 1). Interestingly, at $B = 1$ T, $\Delta T_{fl}$ increases sharply by about 3 times, then unexpectedly decreases by about 1.2 times at $B = 7$ T, and then remains almost independent of $B$ (Table 1). It is assumed that the unexpected sharp ($\approx 2$ times) decrease in $\Delta T_{fl}$ at $B = 0.5$ T is associated with a change in the mechanisms of interaction of the magnetic field with the electronic subsystem of the HTSC, which is reflected as a change in the FLC from 2D–MT to 2D–AL.

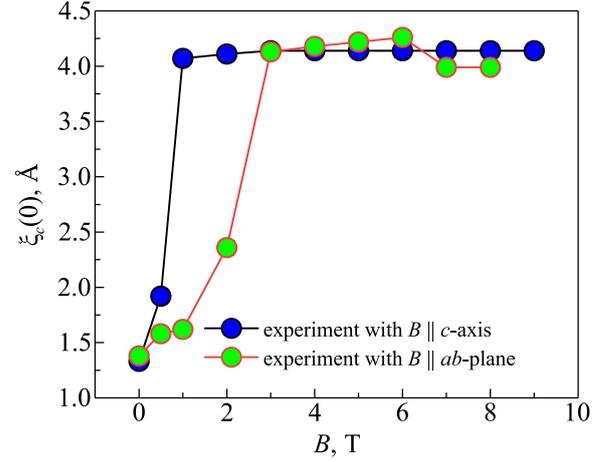

*Fig. 7.* (Color online) The dependence of $\xi_c(0)$ on the magnetic field $B$ is calculated according to Eq. (6), for the cases of $B \parallel c$ (blue dots) and $B \parallel ab$ (green dots) of the $YBa_2Cu_3O_{7-\delta}$ film.

Also, in small fields, a noticeable difference is observed from the results obtained in a field parallel to the $ab$ plane (green dots) [33] (Fig. 7), which reach a maximum value only at $B > 3$ T. When $B \parallel c$, $\xi_c(0)$ reaches its maximum value already in field $B = 1$ T, and then practically does not change with increasing field. It is assumed that for $B \parallel c$, it is necessary to take into account not only the spin-orbit, but also the Zeeman effect [54]. That is, in this case, the interaction of the field with the charge carrier subsystem of the HTSC is assumed to be more effective. Considering that in the classical theory of superconductivity [31, 46]:

Table 1. Parameters of the analysis of the fluctuation conductivity (FLC) of the $YBa_2Cu_3O_{7-\delta}$ film depending on the applied magnetic field

| $B$, T | $T_c$, K | $T_c^{mf}$, K | $T_0$, K | $T_{01}$, K | $T_G$, K | $\Delta T_{fl}$, K | $\xi_c(0)$, Å | $d_{01}$, Å |
|---|---|---|---|---|---|---|---|---|
| 0 | 88.8 | 89.73 | 90.9 | 100.7 | 90.09 | 10.6 | 1.33 | 3.82 |
| 0.5 | 87.7 | 89.23 | 91.6 | 95.2 | 89.73 | 5.5 | 1.92 | 7.42 |
| 1 | 86.7 | 88.62 | 99.4 | 119.9 | 89.48 | 30.4 | 4.07 | 6.85 |
| 2 | 84.8 | 87.99 | 98.9 | 120.0 | 89.31 | 30.7 | 4.11 | 6.82 |
| 3 | 83 | 87.55 | 98.5 | 120.1 | 88.89 | 31.2 | 4.14 | 6.78 |
| 4 | 81.2 | 87.01 | 97.9 | 120.0 | 88.72 | 31.3 | 4.14 | 6.72 |
| 5 | 79.5 | 86.61 | 97.4 | 120.1 | 88.38 | 31.7 | 4.14 | 6.65 |
| 6 | 77.7 | 86.09 | 96.8 | 120.1 | 88.23 | 31.9 | 4.14 | 6.58 |
| 7 | 75.9 | 85.60 | 96.3 | 114.9 | 88.06 | 26.8 | 4.14 | 7.06 |
| 8 | 74.3 | 85.25 | 95.9 | 115.1 | 87.78 | 27.3 | 4.14 | 6.99 |
| 9 | 72.8 | 84.83 | 95.4 | 115.1 | 87.68 | 27.4 | 4.14 | 6.92 |

$$\xi_0 \sim \frac{\hbar v_F}{\pi \Delta(0)}, \qquad (8)$$

that is, $\xi_0 \sim 1/T_c$, since $\Delta = 1.76 \, k_B T_c$ [31], the dependence of $\xi_c(0)$ on $T_c$ was constructed (Fig. 8, blue dots). A noticeable difference is visible between the obtained results and the classical dependence $\xi_0(T_c)$ [Eq. (8)], which is shown by green squares.

Thus, the effect of the magnetic field on the HTSC becomes even more specific than we observed in our previous work [33]. In our case, the field increases $\xi_c(0)$, i.e., the size of the FCPs in the range of SC fluctuations near $T_c$ (Figs. 7 and 8), as well as the region of SC fluctuations $\Delta T_{fl}$ and the distance between conducting planes CuO$_2$, $d_{01}$ — already starting from $B = 1$ T (Table 1). The increase in FCPs seems logical, since $\xi_c(0) \sim 1/T_c$ [Eq. (8)], and $T_c$ decreases by 16 K (approximately 1.22 times) (Table 1). However, $\xi_c(0)$ increases very sharply — almost 3 times already at $B = 1$ T, and then it does not depend on either the field or on $T_c$ (Figs. 7 and 8). It is obvious that the mechanisms of the effect of the magnetic field on FCPs are qualitatively different from those predicted by the theory in the absence of a magnetic field. At first glance, the increase in the size of the pairs can be explained by the fact that the field eventually breaks the pairs, reducing $T_c$ and increasing the magnetoresistance (Fig. 2). At the same time, the increase in $\xi_c(0)$ also leads to the observed expansion of the SC fluctuation region $\Delta T_{fl}$. Less clear is the transition of the temperature dependence of 2D fluctuations from 2D–MT to 2D–AL at $B \geq 1$ T (Fig. 5). Most likely, the magnetic field forms a two-dimensional vortex lattice in the film, which leads to the observation of 2D–AL fluctuations of fluctuating Cooper pairs above the temperature $T_0$, when the field exceeds 1 T. Even less clear is the approximately 1.8-fold increase in the value of $d_{01}$ (Table 1), which follows from the calculations. $d_{01}$ is determined by the simple formula:

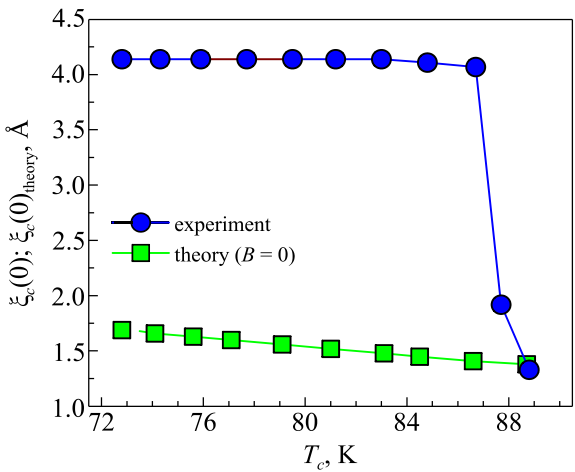

*Fig. 8.* (Color online) Temperature dependences of $\xi_c(0)$ for the YBa$_2$Cu$_3$O$_{7-\delta}$ film at $B \parallel c$ (blue dots) and theoretical values of $\xi_c(0)$ calculated at $B = 0$ using Eq. (8) (green squares).

$d_{01} = d(\varepsilon_0/\varepsilon_{01})^{1/2}$, assuming that at the temperature $T = T_{01}$, when the data finally deviate from the classical theory, the equality $\xi_c(T_{01}) = d_{01}(\varepsilon_{01})^{1/2}$ is satisfied. In this case, we can assume either that this approach no longer works in a magnetic field, and another formula is needed, or that the obtained values of $d_{01}$ correspond, for example, to the period of a two-dimensional vortex lattice formed under the action of the field. It is believed that the analysis of the PG can clarify these issues.

### 3.3. Field analysis of the pseudogap

It is known that in cuprates at temperatures below $T^*$, a decrease in the density of states at the Fermi level is observed [55, 56], which is called a pseudogap [1, 15, 57, 58]. At the same time, $\rho(T)$ deviates downward from linearity, which, as mentioned above, leads to the appearance of excess conductivity [Eq. (1)]. It is believed that $\sigma'(T)$ should contain information about the magnitude and temperature dependence of the PG [59]. In the absence of a rigorous theory of HTSCs, the question of obtaining this information remains uncertain. We hold the view that the PG is associated with the formation of LPs at $T < T^*$ which at $T \leq T^*$ behave as SBBs obeying the Bose–Einstein condensation (BEC) theory, but with decreasing temperature they gradually change their properties to fluctuating Cooper pairs near $T_c$, obeying the Bardeen–Cooper–Schrieffer (BCS) theory [3, 6, 7, 12–15, 59]. It is known that classical fluctuation theories (3D–AL [50] and 2D–MT [48]) describe $\sigma'(T)$ well only in a relatively narrow temperature region (no more than 20 K, above $T_c$) [12, 15, 20, 36, 37]. Obviously, to obtain information about the PG in the entire temperature range from $T^*$ to $T_G$, an equation is needed that would describe the entire experimental curve $\sigma'(T)$ and contain the PG parameter $\Delta^*(T)$ in an explicit form. Such an equation was proposed in [59], taking into account the local pair (LP) model:

$$\sigma'(T) = A_4 \frac{e^2 \left(1 - \frac{T}{T^*}\right) \exp\left(-\frac{\Delta^*}{T}\right)}{16\hbar \xi_c(0) \sqrt{2\varepsilon^*_{c0}} \sinh\left(2\frac{\varepsilon}{\varepsilon^*_{c0}}\right)}, \qquad (9)$$

where $(1 - T/T^*)$ and $\exp(-\Delta^*/T)$ take into account the dynamics of LPs formation at $T \leq T^*$ and their destruction by thermal fluctuations $kT$ near $T_c$, respectively. Solving Eq. (9) with respect to $\Delta^*(T)$, we obtain:

$$\Delta^*(T) = T \ln \frac{e^2 A_4 \left(1 - \frac{T}{T^*}\right)}{\sigma'(T) 16\hbar \xi_c(0) \sqrt{2\varepsilon^*_{c0}} \sinh(2\varepsilon/\varepsilon^*_{c0})}, \qquad (10)$$

where $\sigma'(T)$ is the experimentally measured excess conductivity in the entire temperature range from $T^*$ to $T_G$, $A_4$ is a numerical coefficient (analogous to the $C$-factor in the theory of fluctuation conductivity), $\Delta^*(T_G)$ is the value of the PG parameter near $T_c$, and $\varepsilon^*_{c0}$ is a theoretical parameter [60].

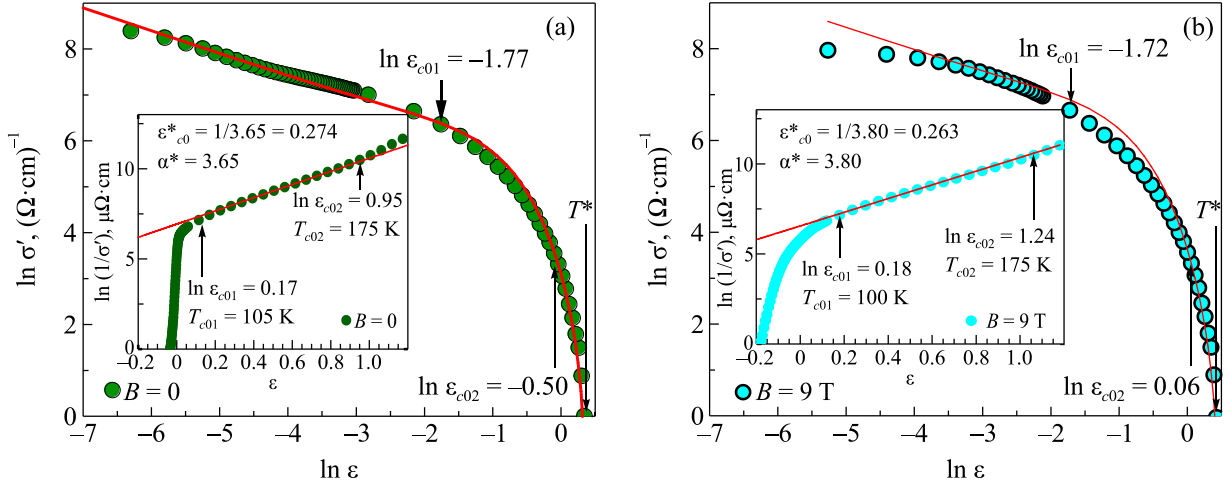

*Fig. 9.* (Color online) ln σ′ vs ln ε in the temperature range from $T^*$ to $T_G$ for (a) $B = 0$ (green dots) and (b) $B = 9$ T (turquoise dots). The red solid curves in each panel are the approximation of the data by Eq. (9). Insets: ln σ′$^{-1}$ as a function of ε. The straight red lines indicate the linear portions of the curves between $\varepsilon_{c01}$ and $\varepsilon_{c02}$.

Parameters such as $T^*$, $T_c^{mf}$, the reduced temperature $\varepsilon = (T - T_c^{mf})/T_c^{mf}$, and $\xi_c(0)$ for different values of $B$ have been defined previously (see Table 1). Recall that $T_c^{mf}$ separates the region of SC fluctuations from the critical fluctuations of the region $T_c$, where the mean field theory does not work [31, 61, 62]. The definition of this temperature was described in Sec. 3.2 (Fig. 4). The Ginzburg temperature $T_G > T_c^{mf}$, down to which the mean field theory works [63, 64], is another characteristic of the proposed PG analysis. All these parameters are given in Table 1 for all values of the magnetic field. Other parameters such as $T_c$, $T_0$, $T_{01}$, and $\Delta T_{fl}$ are also shown in Table 1. As can be seen from Table 1, the magnetic field affects all characteristic temperatures except $T^* = 219$ K. This independence of $T^*$ from $B$ is confirmed by data for YBCO [57] and Bi-2212 [65, 66] and is consistent with the known fact that even a relatively strong magnetic field (≈ 80 T [57]) has no noticeable effect on the resistivity of cuprates in the normal state [39].

It is important that all the missing parameters required for Eq. (9) and Eq. (10), such as $\varepsilon^*_{c0}$, $\Delta^*(T_G)$, and the coefficient $A_4$, can be experimentally determined using the approach developed within the framework of the LP model [15, 16, 59]. Figure 9 shows the dependences of ln σ′ on ln ε for $B = 0$ and $B = 9$ T in the entire temperature interval from $T^*$ to $T_G$. It is theoretically shown that σ′$^{-1}$ ~ exp(ε) in a certain temperature interval, indicated by the arrows ln $\varepsilon_{c01}$ and ln $\varepsilon_{c02}$ in the main panels [60]. This feature turns out to be one of the main properties of most HTSCs. As a result, in the interval from $\varepsilon_{c01}$ to $\varepsilon_{c02}$ (see the insets in the figures), lnσ′$^{-1}$ is a linear function of ε with slope α*, which determines the parameter $\varepsilon^*_{c0} = 1/\alpha^*$. This approach allows us to obtain reliable values of $\varepsilon^*_{c0}$ for all values of the applied magnetic field (Table 2), which significantly affect the shape of the theoretical curves ln σ′ on ln ε at large $T$. Interestingly, starting from $B = 0.5$ T, $\varepsilon^*_{c0} = 0.263$ regardless of the field.

Figure 10 presents the same dependences for $B = 0$ and $B = 9$ T, but in the coordinates of ln σ′ on $1/T$. As established in [12, 15, 16, 52, 53], in this case, the PG parameter $\Delta^*(T_G)$ significantly affects the shape of the theoretical curves presented in Fig. 10, at $T > T_{01}$, i.e., noticeably above the region of SC fluctuations. In addition, it was established that $\Delta^*(T_G) = \Delta(0)$, where $\Delta$ is the SC order parameter [67, 68]. We emphasize that it is the value of $\Delta^*(T_G)$ that determines the true value of the PG and is used to estimate the value of the BCS ratio $D^* = 2\Delta(0)/k_BT_c = 2\Delta^*(T_G)/k_BT_c$ in a specific HTSC sample [15, 16, 53, 69]. The best approximation of $\ln \sigma'$ as a function of $1/T$ from Eq. (9) for the case $B = 0$ is achieved at a value of $D^* = 5.0$ (Table 2) and $D^* = 5.2$ ($B = 9$ T), which corresponds to the strong binding limit characteristic of YBCO [70–72].

Table 2. Parameters of the PG analysis of the $YBa_2Cu_3O_{7-\delta}$ film depending on the applied magnetic field

| $B$, T | $T_{pair}$, K | $\varepsilon^*_{c0}$ | $C_{3D}$ | $A_4$ | $D^*$ | $\Delta^*(T_G)/k_B$, K | $\Delta^*(T_{pair})/k_B$, K |
|---|---|---|---|---|---|---|---|
| 0 | 140 | 0.271 | 0.39 | 7.8 | 5.0 | 223.4 | 254.4 |
| 0.5 | 135 | 0.263 | 0.68 | 12.3 | 5.0 | 220.3 | 257.3 |
| 1 | 135 | 0.263 | 1.6 | 27.5 | 4.9 | 209.1 | 259.3 |
| 2 | 135 | 0.263 | 1.69 | 28.1 | 4.95 | 208.7 | 255.3 |
| 3 | 133 | 0.263 | 1.77 | 29.2 | 5 | 206.0 | 255.9 |
| 4 | 130 | 0.263 | 1.8 | 29.8 | 5 | 205.1 | 253.7 |
| 5 | 130 | 0.263 | 1.89 | 30.2 | 5 | 203.0 | 252.1 |
| 6 | 130 | 0.263 | 1.95 | 31.7 | 5.1 | 202.7 | 253.9 |
| 7 | 130 | 0.263 | 2.03 | 32.0 | 5.1 | 199.0 | 250.9 |
| 8 | 130 | 0.263 | 2.08 | 32.5 | 5.2 | 198.9 | 249.8 |
| 9 | 130 | 0.263 | 2.16 | 32.0 | 5.2 | 194.2 | 244.0 |

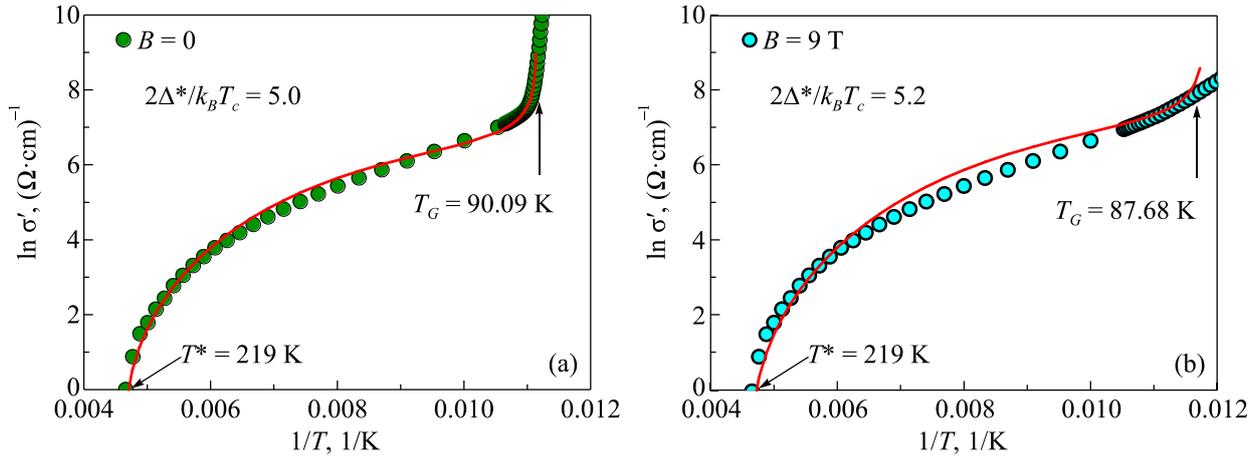

*Fig. 10.* (Color online) $\ln \sigma'$ vs $1/T$ for $B = 0$ [(a), green dots] and $B = 9$ T [(b), turquoise dots] over the entire temperature range from $T^*$ to $T_G$. The red solid curves in each panel are approximations of the data using Eq. (9). The best approximation is observed at $2\Delta^*(T_G)/k_BT_c = 5.0$ ($B = 0$) and $2\Delta^*(T_G)/k_BT_c = 5.2$ ($B = 9$ T).

After determining the parameters $\varepsilon^*_{c0}$, $\Delta^*(T_G)$, the coefficient $A_4$ in Eq. (9) was found by approximating the experimental data $\ln \sigma'$ vs $\ln \varepsilon$ by Eq. (9) for each value of the magnetic field (red curves in Fig. 9). Previously established parameters from Tables 1 and 2 were also used. For the correct choice of $A_4$, the theory is combined with the experiment in the region of 3D fluctuations near $T_c$, where the dependence $\ln \sigma'(\varepsilon)$ vs $\ln \varepsilon$ is a linear function with slope $\lambda = -1/2$ (Fig. 9). It is important to emphasize that the obtained Eq. (9) (solid red curves), taking into account the full set of determined parameters, demonstrates an excellent ability to describe the behavior of excess conductivity $\sigma'(\varepsilon)$ in the entire temperature range from $T^*$ to $T_G$ both at $B = 0$ and for all applied magnetic fields, which confirms the universality of the approach. But it should be noted that despite the general good agreement of theory with experiment, in the range of magnetic fields above 2 T, systematic small but noticeable deviations of the experimental values of $\sigma'(T)$ from the theoretical curve constructed by Eq. (9) were found at $T_{01} > T > T_0$ due to the change of 2D fluctuations from 2D–MT to 2D–AL. We note that similar deviations of the theory [Eq. (9)] from the experiment in the range of 2D fluctuations, are also clearly observed in Figs. 9 and 10 at $B = 9$ T.

It is important to note that all model parameters (see Table 2), except for $\varepsilon^*_{c0}$, demonstrate a clear dependence on the magnitude of the applied magnetic field. The BCS ratio, $D^* = 2\Delta^*(T_G)/k_BT_c$, which characterizes the strength of the superconducting coupling, at $B > 0.5$ T tends to decrease slightly (from 5.0 at $B = 0$ to 4.9 at $B = 1$ T, Table 2). However, with increasing magnetic field ($B > 2$ T), this value stabilizes at $\approx 5.0$ ($B = 2$–5 T), and then shows a gradual increase to 5.2 at $B = 9$ T. But the scaling coefficient $A_4$ changes most significantly (Table 2). Note that at $B = 0$, $A_4$ has the smallest value of 7.8, which is typical for YBCO films with a good structure [59]. But with increasing field $A_4$, it also increases rapidly to 27.5 at $B = 1$ T. Then it slowly increases to 30.2 at $B = 5$ T, and then remains at $32 \pm 0.5$ at $B = (6$–$9)$ T (Table 2). By the way, the $C_{3D}$ factor also increases (Table 2). Recall that the introduction of scaling factors ($C_{3D}$, $A_4$, Table 2) is due to the inhomogeneity of the current distribution across the sample, which is caused by defects, and which cannot be taken into account theoretically [37]. This allows us to conclude that the magnetic field somehow still affects the inhomogeneity of the sample. If not directly by creating structural defects, then, perhaps, by generating vortices, which can be additional centers of charge carriers scattering, inhomogeneously distributed over the volume of the sample. This clearly indicates that the magnetic field is not a neutral factor for the system and should significantly affect the pseudogap, which is reflected in the change of its parameters (Table 2 and Fig. 11). Our analysis using Eq. (10) showed that it is this combined change in the film parameters under the action of a magnetic field that is the cause of the unexpected behavior of $\Delta^*(T)$ in strong fields ($B = 9$ T) (Fig. 11), despite the fact that the excess conductivity $\sigma'(T)$ changes relatively little with increasing field (Figs. 5, 6, 8, 9).

In the absence of a magnetic field, the temperature dependence $\Delta^*(T)$ (green dots in Fig. 11) was obtained with the following set of experimentally found parameters: $T_c^{mf} = 89.73$ K, $T^* = 219$ K, $\xi_c(0) = 1.33$ Å, $\varepsilon^*_{c0} = 0.271$, $A_4 = 7.8$, and $\Delta^*(T_G)/k_B = 223.4$ K (Tables 1 and 2). The figure also shows the dependences $\Delta^*(T)$ for $B = 3$ T (magenta dots) and $B = 9$ T (turquoise dots), calculated using the corresponding sets of found parameters. The dependence $\Delta^*(T)$ at $B = 0$ fully corresponds to the known behavior for high-quality YBCO films [15, 53, 59] and untwinned single crystals [73]. It is characterized by a broad maximum at $T_{pair} = 140$ K, which is interpreted as a crossover point between the Bose–Einstein condensation and Bardeen–Cooper–Schrieffer theories [8, 10–14, 59], and a minimum at a temperature of $T_{01}$ [15, 49, 53, 59, 73]. With further approach to the critical temperature $T_c$ (as shown in detail in Fig. 12), another local (fluctuation) maximum $\Delta^*(T)$ is observed slightly below the 3D–2D crossover temperature $T_0$, and then a small minimum at the Ginzburg temperature $T_G$, which is typical for defect-free HTSCs [15, 53, 73]. Below $T_G$, the values of $\Delta^*(T)$ are deliberately not given in the graphs, since in the region of critical fluctuations, the LP model does not work [33, 53, 59]. This approach allowed us to accurately determine the value of $T_G$ for each field and, accordingly, to reliably determine the value of the pseudogap $\Delta^*(T_G)/k_B$ at $T = T_G$ [15, 53, 67, 68] (Table 2 and Figs. 12 and 13).

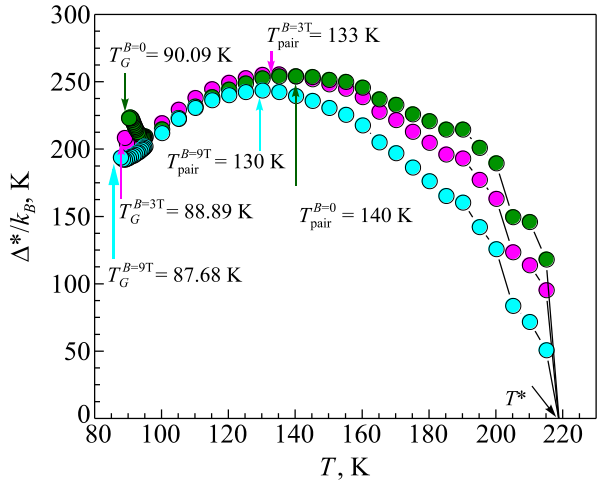

*Fig. 11.* (Color online) Pseudogap $\Delta^*$ as a function of temperature $T$ for the $YBa_2Cu_3O_{7-\delta}$ film, calculated using Eq. (10), with parameters specified in the text, for $B = 0$ (green dots), 3 T (magenta dots), and 9 T (turquoise dots).

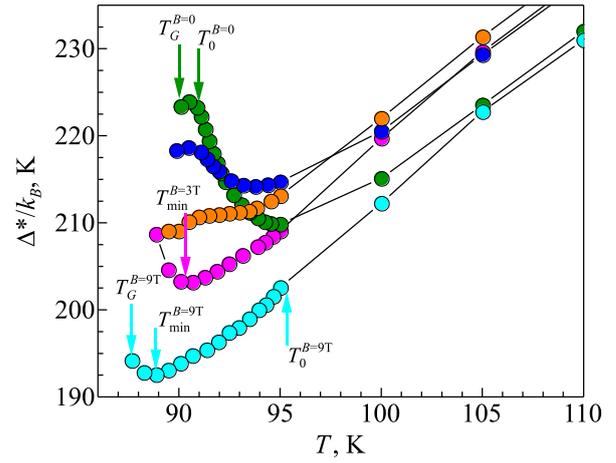

*Fig. 12.* (Color online) The same dependences as in Fig. 11 for the temperature interval $T_G < T < T_{01}$ at $B = 0$ (green dots), 0.5 (blue dots), 1 (orange dots), 3 (magenta dots), and 9 T (turquoise dots). All characteristic temperatures are indicated by arrows for $B = 0$ and 9 T. In both drawings, the thin, solid curves are guides for the eyes.

It is clearly seen from Fig. 11 that the magnetic field significantly modifies the value of $\Delta^*(T)$. The most striking effect is observed at high temperatures ($T > T_{pair}$) for fields $B \geq 3$ T. The values of $\Delta^*(T)$ decrease noticeably in the entire temperature range above $T_{pair}$ compared to $B = 0$, which is especially clearly visible in the field $B = 9$ T. This result is extremely important and somewhat unexpected. Because, according to Fig. 2 and the literature data [37, 54], the magnetic field has practically no effect on the resistivity $\rho(T)$ in the normal state. This means that our method of analyzing the excess conductivity $\sigma'(T)$ and the PG parameter $\Delta^*(T)$ exhibits exceptional sensitivity to the effect of the magnetic field on the subsystem of charge carriers associated with the formation of local pairs and a pseudogap. This allows us to draw a key conclusion that the magnetic field has a stronger effect on the state and dynamics of local pairs (which form the PG) than on normal (unpaired) conduction electrons. In addition, it is necessary to note the nontrivial and unexpected dependence of $\Delta^*(T)$ in the temperature range $T < T_{pair}$ in fields from 1 to 3 T, when $\Delta^*(T)$ passes above the data obtained at $B = 0$. At $B = 3$ T, it seems that the entire curve $\Delta^*(T)$ undergoes a shift towards lower temperatures when a magnetic field is applied. In this case, the temperature $T_{pair}$ decreases with increasing field (from 140 K at $B = 0$ to 130 K at $B = 9$ T, Table 2). At the same time, the absolute value of $\Delta^*(T_{pair})/k_B$ initially increases slightly to 255.9 K at $B = 3$ T, but then gradually decreases to 244.0 K at $B = 9$ T (Table 2). Note that the same dependence $\Delta^*(T)$ as at $B = 3$ T is observed for the case of a field directed perpendicular to the $c$ axis [32], but at $B = 8$ T. Interestingly, the value of $\Delta^*(T_{pair})/k_B$ did not decrease [32]. This indicates a more effective suppression of the superconducting properties of the sample under study under the influence of a magnetic field directed along the $c$ axis, when, as noted above, both spin-orbital and Zeeman effects must be taken into account [54, 66]. As a result, as can be seen from Fig. 11, at $B = 9$ T, the decrease in $\Delta^*(T)$ becomes even more significant and extends to the entire temperature range under study.

The question arises: how to explain this very important and somewhat unexpected result? Analysis of the obtained data shows that of all the parameters included in Eq. (10) for the PG, only the coherence length $\xi_c(0)$ (Table 1 and Figs. 7 and 8) changes noticeably with increasing magnetic field, which also determines the size of local pairs. This, in turn, leads to an increase in the scaling factors $C_{3D}$ and $A_4$, which, together with $\xi_c(0)$, are included in Eqs. (3) and (10). Already at $B = 1$ T, $\xi_c(0)$ increases more than threefold, which, undoubtedly, should cause the observed decrease in $\Delta^*(T)$. However, the following question arises: why does $\Delta^*(T)$ continue to decrease with increasing field, while at $B > 1$ T, $\xi_c(0)$ remains practically unchanged (see Table 1)? Moreover, if the increase in $d_{01}$ (Table 1) may raise doubts, then the evolution of $\xi_c(0)$ under the action of the field is an experimental fact, since it is determined by the temperature of the 3D–2D crossover within the classical AL and MT fluctuation theories. Therefore, the change in $\xi_c(0)$ is clearly insufficient to explain the decrease in $\Delta^*(T)$ in the magnetic field. Apparently, it should be mentioned that by definition, the PG is a decrease in the density of states at the Fermi level [18, 20, 24, 25]. It can be assumed that the unusual behavior of $\Delta^*(T)$ in the high-temperature region is due to the specific interaction of the magnetic field with local pairs in HTSCs, which, as is known [13–15], at $T \leq T^*$ exist in the form of strongly bound bosons (SBBs) obeying the BEC. This approach also allows us to explain the growth of $\Delta^*(T)$ observed at $T < T_{pair}$ in weak fields. Indeed, at $T < T_{pair}$, SBBs are transformed into already formed Cooper pairs [14, 46], the mechanism of interaction of the magnetic field with which may be different. Obviously, for the final solution of this issue, it is necessary to develop an adequate theory that would take into account the mechanisms of interaction of the magnetic field with charge carriers in HTSC in the region of existence of the PG.

As can be seen from Fig. 11, when the temperature approaches $T_c$ (values at different $B$ in Table 1), the magnetic field has an intense effect on $\Delta^*(T)$ precisely in the region of superconducting fluctuations, i.e., below ($T_{01} = 100.7$ K at $B = 0$). The sharp low-temperature fluctuation maximum observed near $T_0$ at $B = 0$ is gradually suppressed in amplitude, shifts towards higher temperatures, and completely disappears at magnetic fields $B \approx 1$ T (orange dots). However, starting from $B > 1$ T, a clearly pronounced minimum appears on $\Delta^*(T)$ at $T_{min} > T_G$. As can be seen from the figure, $\Delta^*(T)$, which corresponds to the minimum, noticeably decreases with increasing magnetic field. Accordingly, $\Delta^*(T_G)$, which is the leftmost point of each curve and at the same time determines $T_G$, also decreases and shifts towards lower temperatures.

The observed non-trivial decrease in $\Delta^*(T)$ in the region of SC fluctuations near $T_c$ is very likely due to two main factors. First, it is the destruction of local pairs by a magnetic field, which is considered the main mechanism of the appearance of magnetoresistance and the decrease in $T_c$ [32, 33]. Second, it is assumed that at $B > 1$ T a two-dimensional vortex lattice is formed in the film [33] (and references therein), which can modify the mechanisms of interaction of the magnetic field with charge carriers and affect the PG. The decrease in $\Delta^*(T_G)$ with increasing $B$ shown in Fig. 13 looks very interesting and unusual and may be an additional manifestation of this modification. Indeed, we can assume the following scenario of the influence of a magnetic field on a YBCO film in the region of SC fluctuations. Recall that at $T < T_{01}$, local pairs are transformed into fluctuating Cooper pairs (FCPs) [13–15], which, in fact, leads to the transition of the HTSCs to the SC state. It is obvious that the magnetic field effectively destroys the SC fluctuations. Accordingly, already at $B = 1$ T the fluctuation maximum at $T = T_0$ completely disappears (orange dots in Fig. 11). At the same time, as can be seen from Fig. 13, in the field interval $0 < B < 1$ T, $\Delta^*(T)$ rapidly decreases as $d\Delta^*/dB \approx$ $\approx -14.3$ K/T. Most likely, this occurs due to the destruction of the FCPs by the magnetic field, which, most likely, can be described within the framework of the Abrikosov–Gorkov pair-breaking theory modified for $d$-wave superconductors [74]. It is most likely that after this, a two-dimensional vortex lattice is formed in the HTSC, the interaction of which with the magnetic field now determines the behavior of $\Delta^*(T)$. It is clearly seen (Fig. 12) that for $B \geq 1$ T the nature of this interaction changes noticeably. At $B > 1$ T, a broad "magnetic-vortex" minimum appears on the $\Delta^*(T)$ dependence at $T_{min}$ (Fig. 12). After this, $\Delta^*(T_G)$ continues to decrease linearly, but with a completely different, much lower intensity $d\Delta^*/dB \approx -1.86$ K/T (Fig. 13). At the same time, the Ginzburg temperature also demonstrates a peculiarity at $B > 1$ T, as shown in Fig. 14.

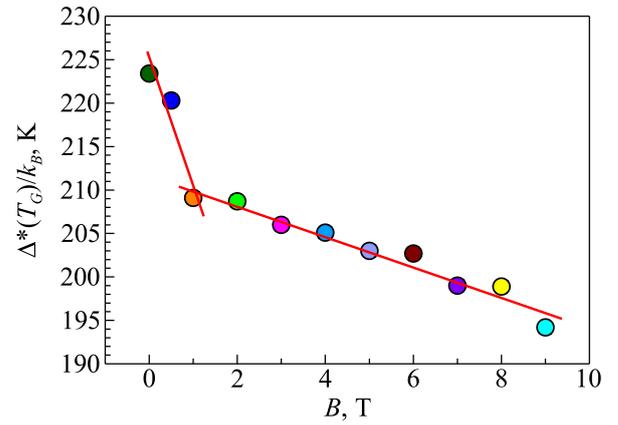

*Fig. 13.* (Color online) Pseudogap $\Delta^*(T_G)$ as a function of magnetic field $B$ for YBa$_2$Cu$_3$O$_{7-\delta}$ film.

Within the framework of this approach, it was logical to compare the dependence $\Delta^*(T_G)$ with similar results obtained for the case when $B \parallel ab$ (see Fig. 6 in [32]). As can be seen from the figure, two sections are also observed on the dependence $\Delta^*(T_G)$. In the field interval $0 < B < 2$ T, $\Delta^*(T_G)$ decreases with intensity: $d\Delta^*/dB = -1.94$ K/T, which is more than 7 times less than in the field $B \parallel c$. Accordingly, the fluctuation maximum completely disappears only at $B > 5$ T. However, at $B = 2$ T, a parallel shift of all data by 1 T to the right is observed, which is logically associated with the beginning of the formation of a vortex lattice in the film. Based on this comparison, we can conclude that in the region of small fields, a magnetic field directed along the $c$ axis, where both spin-orbital and Zeeman effects must be taken into account, destroys the FCPs much more effectively and suppresses SC fluctuations already at $B = 1$ T. However, the intensity of the interaction with a two-dimensional vortex lattice in the region of large fields in both cases is approximately the same: $d\Delta^*/dB = (-1.9 \pm 0.04)$ K/T.

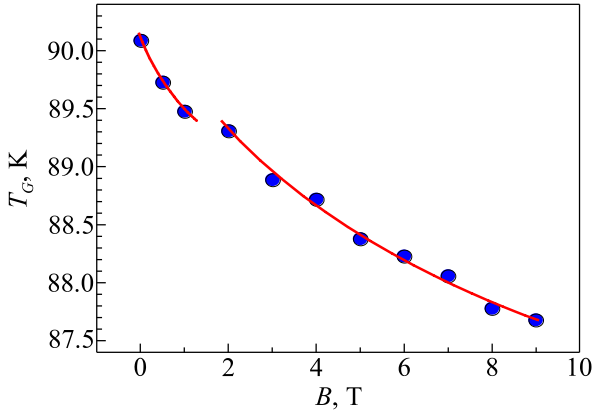
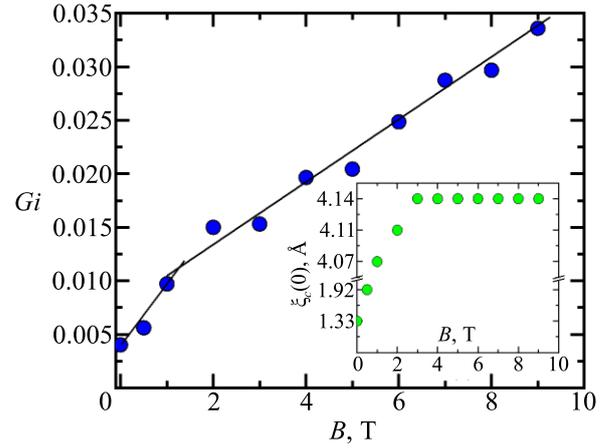

*Fig. 14.* (Color online) Ginzburg temperature $T_G$ as a function of magnetic field $B$. In both drawings, the solid lines and curves are guides for the eyes.

*Fig. 15.* (Color online) Dependence of the Ginzburg parameter $Gi$ on the magnetic field $B$ (blue dots). The solid lines are guides for the eyes. The inset shows $\xi_c(0)$ as a function of $B$ on a reduced scale.

Additional interesting information can be obtained by analyzing how the magnetic field affects the size of the critical fluctuation region, which is determined by the Ginzburg parameter: $Gi = (T_G - T_c^{mf})/T_c^{mf}$ [63, 75, 76]. It follows from Fig. 4 and Table 1 that the magnetic field significantly increases the difference $T_G - T_c^{mf}$. As a result, the Ginzburg parameter also increases (see Fig. 15), and accordingly: $Gi(0) = 0.004$ and $Gi(9\,\text{T}) = 0.034$ (Table 3). Moreover, just like $\xi_c(0)$ (Fig. 7) and $\Delta^*(T_G)$ (Fig. 13), $Gi(B)$ grows non-monotonically (Fig. 15). In the field range $0 < B < 1$ T, the parameter $Gi$ grows with an intensity $dGi/dB = 5.5 \cdot 10^{-3}$ T$^{-1}$, and in the range $2 < B < 9$ T — with a much lower intensity: $dGi/dB = 3.0 \cdot 10^{-3}$ T$^{-1}$. It is important that the change in the nature of the dependences (marked by solid lines in Fig. 15) occurs in the same field interval (1–2 T), where the magnetic field completely suppresses SC fluctuations, and a vortex lattice is formed in HTSC with a high probability. The inset shows the evolution of $\xi_c(0)$ on a smaller scale. It is evident that in the interval $0.5\,\text{T} < B < 1$ T, $\xi_c(0)$ demonstrates a jump from 1.92 to 4.07 Å (see also Figs. 7 and 8). After that, there follows a transition region $1\,\text{T} < B < 3$ T. And only at $B \geq 3$ T, $\xi_c(0)$ reaches a constant value of 4.14 Å (Tables 1 and 3). Thus, the idea of the possibility of forming a two-dimensional vortex lattice in a YBCO film under the action of a magnetic field in the region of the SC transition finds additional confirmation in the field dependence of the Ginzburg parameter $Gi(B)$.

It should be emphasized that the dependence $Gi(B)$ obtained in a magnetic field (Fig. 15) is fundamentally different from the dependence $Gi(P)$ obtained, for example, under the action of hydrostatic pressure $P$ on cuprate HTSCs. In this case, naturally, no vortex lattices arise, and $Gi(P)$ grows smoothly, demonstrating a parabolic dependence on $P$ [75, 76].

According to the anisotropic Ginzburg–Landau theory, the Ginzburg number is given as [63, 64]:

$$Gi = \alpha\left(\frac{k_B}{\Delta c \xi_c(0) \xi_{ab}^2(0)}\right)^2, \qquad (11)$$

Table 3. Parameters for analyzing the $Gi$ criterion

| $B$, T | $T_c$, K | $T_G$, K | $T_c^{mf}$, K | $C_{3D}$ | $\xi_c(0)$, Å | $Gi$ |
|---|---|---|---|---|---|---|
| 0 | 88.8 | 90.1 | 89,73 | 0.39 | 1.33 | 0.00401 |
| 0.5 | 87.7 | 89.7 | 89.23 | 0.68 | 1.92 | 0.0056 |
| 1 | 86.7 | 89.5 | 88,62 | 1.6 | 4.07 | 0.0097 |
| 2 | 84.8 | 89.3 | 87,99 | 1.69 | 4.11 | 0.015 |
| 3 | 83 | 88.9 | 87,55 | 1.77 | 4.14 | 0.0153 |
| 4 | 81.2 | 88.7 | 87,01 | 1.8 | 4.14 | 0.0197 |
| 5 | 79.5 | 88.4 | 86,61 | 1.89 | 4.14 | 0.0204 |
| 6 | 77.7 | 88.2 | 86,09 | 1.95 | 4.14 | 0.0249 |
| 7 | 75.9 | 88.1 | 85,60 | 2.03 | 4.14 | 0.0287 |
| 8 | 74.3 | 87.8 | 85,25 | 2.08 | 4.14 | 0.0297 |
| 9 | 72.8 | 87.7 | 84,83 | 2.16 | 4.14 | 0.0336 |

where α is a constant of the order $10^{-3}$, $\Delta c$ is the jump of the specific heat at $T_c$, and $\xi_{ab}(0)$ is the in-plane coherence length. According to microscopic theory [64], $\Delta c \sim T_c N(0)$, where $N(0)$ is the single-particle density of states at the Fermi level. The question arises: what can cause $Gi$ to increase (Fig. 15) if $\xi_c(0)$ ($B = 9$ T) increases by 3.11 times? Most likely, due to a decrease in $\xi_{ab}(0)$. To make the appropriate estimates, consider the ratio $Gi^* = Gi(9\ \text{T})/Gi(0) \approx 8.4$. It follows that to obtain such a value of $Gi^*$, the condition must be met: $\xi_{ab}(9\ \text{T}) = 1/3\ \xi_{ab}(0)$, i.e., $\xi_{ab}(0)$ must decrease three times. It is generally accepted that $\xi_{ab}(0) \approx 10\ \xi_c(0) \approx 13.3$ Å. As a result, we obtain that $\xi_{ab}(9\ \text{T}) = 13.3/3 \approx 4.43$ Å. As a result, we have a very interesting and unexpected conclusion: in a magnetic field $\xi_c(0) \approx \xi_{ab}(0)$. That is, the pairs become isotropic. If we recall that the magnetic field penetrates a second-type superconductor in the form of Abrikosov vortices, and that Abrikosov created the theory for classical isotropic superconductors, then this result seems quite reasonable. If we assume that the magnetic field reduces the jump in heat capacity $\Delta c$, for example, by half, then we obtain: $\xi_{ab}(9\ \text{T}) = 13.3/2.14 \approx 6.21$ Å. That is, there is still a noticeable decrease in $\xi_{ab}(9\ \text{T})$ with a distinct tendency towards isotropization.

### 4. Conclusion

The work investigates the effect of an external magnetic field parallel to the $c$ axis on the temperature dependence of the fluctuation conductivity (FLC) and pseudogap (PG) in a YBa$_2$Cu$_3$O$_{7-\delta}$ (YBCO) film with a critical temperature $T_c = 88.8$ K (at $B = 0$) within the framework of the local pair (LP) model.

Strictly speaking, the effect of a magnetic field on HTSCs is significantly different from the effect of other external factors, such as irradiation with high-energy electrons [77] or the presence of magnetic impurities [53]. In the latter case, external factors strongly change the resistivity curves of samples and the temperature dependences of the PG. At the same time, a magnetic field, at least up to 20 T, does not affect the resistivity curve of HTSCs (Fig. 2) [37, 54], and therefore, it would seem, should not affect the PG. The effect begins in the region of superconducting (SC) fluctuations at temperatures $T < T_{01}$ (Fig. 2). This leads to a noticeable linear decrease in the critical temperature $T_c$ (Fig. 3) and an increase in the magnetoresistance $MR = [\rho(B) - \rho(0)]/\rho(0)$, probably due to the destruction of fluctuating Cooper pairs (FCPs). Note that, unlike the field directed along the $ab$ plane of the HTSC, $T_c$ decreases by 10 K more. No other information about the internal state of the HTSC can be obtained from resistive measurements. The situation is different with the study of the FLC (Figs. 4–6) and the PG (Figs. 9–14).

It is shown that at $B > 1$ T all parameters of the HTSC change their behavior under the influence of the field: $\xi_c(0)$ increases more than 3 times and then retains a constant value of $\approx 4.14$ Å up to $B = 9$ T (Figs. 7–8). At the same time, the PG in the high-temperature region decreases noticeably — most likely, it is precisely because of this increase in $\xi_c(0)$, the nature of which, strictly speaking, is not fully understood. However, the most interesting behavior is observed in the region of SC fluctuations, i.e., in the temperature interval $T_{01} > T > T_G$, as shown in detail in Fig. 12. First, the magnetic field suppresses SC fluctuations by breaking the FCPs, and already at $B = 1$ T the fluctuation maximum at $T \approx T_0$ completely disappears. At $B > 2$ T, a broad minimum appears, which increases with increasing field. At the same time, the temperature $T_G$, which corresponds to the leftmost point of each graph, shifts towards lower temperatures (Fig. 14). All data indicate that at $B > 1$ T, the mechanisms of interaction of the magnetic field with the HTSC change. This is well illustrated by the field dependence $\Delta^*(T_G)/k_B$ (Fig. 13). In the field range $0 < B < 1$ T, $\Delta^*(T_G)/k_B$ rapidly decreases with a gradient $d\Delta^*/dB \approx -14.3$ K/T. Most likely, this occurs due to the destruction of the FCP by the magnetic field, which can probably be described within the framework of the Abrikosov–Gorkov theory. It is probably after this that a two-dimensional vortex lattice is formed in the HTSC, the interaction of which with the magnetic field determines the further behavior of $\Delta^*(T_G)/k_B$. It is clearly seen (Fig. 13) that for $B \geq 1$ T the nature of this interaction changes significantly. At $B > 1$ T, a broad "magnetic-vortex" minimum appears on the $\Delta^*(T)$ dependence at $T_{\min}$ (Fig. 12). After that, $\Delta^*(T_G)/k_B$ continues to decrease linearly, but with a much lower intensity: $d\Delta^*/dB \approx -1.86$ K/T (Fig. 13).

These conclusions are confirmed by the analysis of the dependence of the Ginzburg parameter $Gi = (T_G - T_c^{mf})/T_c^{mf}$ on the magnetic field (Fig. 15). It is seen that $Gi(B)$ increases non-monotonically. In the field range $0 < B < 1$ T, $Gi$ increases with an intensity $dGi/dB = 5.5 \cdot 10^{-3}$ T$^{-1}$, and in the interval $3\ \text{T} < B < 9\ \text{T}$ — with a much lower intensity: $dGi/dB = 3.0 \cdot 10^{-3}$ T$^{-1}$. It is significant that the change in the dependences, which are marked by solid lines in Fig. 15, occurs in the same field range (1–2 T), where the magnetic field completely suppresses the SC fluctuations, and a vortex lattice is formed in the HTSC with a high probability. Thus, the idea of the possibility of the formation of a two-dimensional vortex lattice in the YBCO film by a magnetic field in the region of the SC transition is further confirmed by the dependence of the Ginzburg parameter $Gi(B)$ on the magnetic field. Looking at Eq. (11), the question arises: due to what can $Gi$ increase (Fig. 15), if $\xi_c(0)$ (at $B = 9$ T) increases by more than 3 times? Most likely — due to a decrease in $\xi_{ab}(0)$. To make the appropriate estimates, the ratio $Gi^* = Gi(9\ \text{T})/Gi(0) \approx 8.4$ was considered. Simple algebra allows us to show that to ensure such a value of $Gi^*$, the condition must be fulfilled: $\xi_{ab}(9\ \text{T}) = 1/3\ \xi_{ab}(0)$, i.e., $\xi_{ab}(0)$ must decrease by 3 times.

It is generally accepted that $\xi_{ab}(0) \approx 10\ \xi_c(0) \approx 13.3$ Å. As a result, we obtain: $\xi_{ab}(9\ \text{T}) = 13.3/3 \approx 4.4$ Å. As a result, a very interesting and unexpected conclusion is obtained, namely: in a magnetic field $\xi_c(9\ \text{T}) \approx \xi_{ab}(9\ \text{T})$, i.e., Cooper pairs become practically isotropic. If we recall that the

*A. S. Kolisnyk, M. V. Shytov, E. V. Petrenko et al.*

magnetic field penetrates a second-order superconductor in the form of Abrikosov vortices, and Abrikosov created a theory for classical isotropic superconductors, then this result seems quite logical. We believe that the results obtained by us can serve as the basis for creating new theoretical models designed to describe the specifics of the behavior of the pseudogap in HTSCs under the action of a magnetic field and confirm the conclusions made.

**Acknowledgments**

The work was supported by the National Academy of Sciences of Ukraine within the scientific program F19-5, as well as grants from the Finnish Academy Nos. 343309 and 367561. One of us (A. L. S.) thanks the Division of Low Temperatures and Superconductivity, INTiBS Wroclaw, Poland, as well as the Department of Physics, LUT University, Lappeenranta, Finland, for their hospitality.

1. A. A. Kordyuk, *Pseudogap from ARPES experiment: three gaps in cuprates and topological superconductivity*, *Low Temp. Phys.* **41**, 319 (2015) [*Fiz. Nizk. Temp.* **41**, 417 (2015)].
2. J. Gao, J. W. Park, K. Kim, S. K. Song, H. R. Park, J. Lee, J. Park, F. Chen, X. Luo, Y. Sun, and H. W. Yeom, *Pseudogap and weak multifractality in 2D disordered Mott charge-density-wave insulator*, *Nano. Lett.* **20**, 6299 (2020).
3. J. L. Tallon, J. G. Storey, J. R. Cooper, and J. W. Loram, *Locating the pseudogap closing point in cuprate superconductors: Absence of entrant or reentrant behavior*, *Phys. Rev. B* **101**, 174512 (2020).
4. Y. Y. Peng, R. Fumagalli, Y. Ding, M. Minola, S. Caprara, D. Betto, M. Bluschke, G. M. De Luca, K. Kummer, E. Lefrançis, M. Salluzzo, H. Suzuki, M. Le Tacon, X. J. Zhou, N. B. Brookes, B. Keimer, L. Braicovich, M. Grilli, and G. Ghiringhelli, *Re-entrant charge order in overdoped* $(Bi,Pb)_{2.12}Sr_{1.88}CuO_{6+\delta}$ *outside the pseudogap regime*, *Nat. Mater.* **17**, 697 (2018).
5. D. Chakraborty, M. Grandadam, M. Y. Hamidian, J. C. S. Davis, Y. Sidis, and C. Pépin, *Fractionalized pair density wave in the pseudogap phase of cuprate superconductors*, *Phys. Rev. B* **100**, 224511 (2019).
6. I. Esterlis, S. A. Kivelson, and D. J. Scalapino, *Pseudogap crossover in the electron-phonon system*, *Phys. Rev. B* **99**, 174516 (2019).
7. G. Yu, D.-D. Xia, D. Pelc, R.-H. He, N.-H. Kaneko, T. Sasagawa, Y. Li, X. Zhao, N. Barisic, A. Shekhter, and M. Greven, *Universal precursor of superconductivity in the cuprates*, *Phys. Rev. B* **99**, 214502 (2019).
8. V. M. Loktev, R. M. Quick, and S. G. Sharapov, *Phase fluctuations and pseudogap phenomena*, *Phys. Rep.* **349**, 1 (2001).
9. R. Haussmann, *Properties of a Fermi liquid at the superfluid transition in the crossover region between BCS superconductivity and Bose–Einstein condensation*, *Phys. Rev. B* **49**, 12975 (1994).
10. O. Tchernyshyov, *Non-interacting Cooper pairs inside a pseudogap*, *Phys. Rev. B* **56**, 3372 (1997).
11. J. R. Engelbrecht, A. Nazarenko, M. Randeria, and E. Dagotto, *Pseudogap above $T_c$ in a model with $d_{x^2-y^2}$ pairing*, *Phys. Rev. B* **57**, 13406 (1998).
12. A. L. Solovjov, *Pseudogap and local pairs in high-$T_c$ superconductors, in: Superconductors-Materials*, *Properties and Applications*, A. Gabovich (ed.), InTech, Rijeka (2012), Chap. 7, p. 137.
13. V. J. Emery and S. A. Kivelson, *Importance of phase fluctuations in superconductors with small superfluid density*, *Nature* **374**, 434 (1995).
14. M. Randeria, *Pre-pairing for condensation*, *Nat. Phys.* **6**, 561 (2010).
15. A. L. Solovjov and K. Rogacki, *Local pairs in high-temperature superconductors: The concept of pseudogap*, *Low Temp. Phys.* **49**, 345 (2023) [*Fiz. Nyzk. Temp.* **49**, 375 (2023)].
16. A. L. Solovjov and V. M. Dmitriev, *Fluctuation conductivity and pseudogap in* YBCO *high-temperature superconductors*, *Low Temp. Phys.* **35**, 169 (2009) [*Fiz. Nizk. Temp.* **35**, 227 (2009)].
17. B. P. Stojković and D. Pines, *Theory of the longitudinal and Hall conductivities of the cuprate superconductors*, *Phys. Rev. B* **55**, 8576 (1997).
18. S. Badoux, W. Tabis, F. Laliberté, G. Grissonnanche, B. Vignolle, D. Vignolles, J. Béard, D. A. Bonn, W. N. Hardy, R. Liang, N. Doiron-Leyraud, Louis Taillefer, and C. Proust, *Change of carrier density at the pseudogap critical point of a cuprate superconductor*, *Nature* (London) **531**, 210 (2016).
19. A. M. Gabovich and A. I. Voitenko, *Spatial distribution of superconducting and charge-density-wave order parameters in cuprates and its influence on the quasiparticle tunnel current*, *Low Temp. Phys.* **42**, 863 (2016) [*Fiz. Nizk. Temp.* **42**, 1103 (2016)].
20. L. Taillefer, *Scattering and pairing in cuprate superconductors*, *Annu. Rev. Condens. Matter Phys.* **1**, 51 (2010).
21. S. Dzhumanov and U. T. Kurbanov, *The coexisting of insulating and metallic/superconducting phases and their competing effects in various underdoped cuprates*, *Modern Phys. Lett. B* **32**, 1850312 (2018).
22. S. Dzhumanov and U. T. Kurbanov, *The new metal–insulator transitions and nanoscale phase separation in doped cuprates*, *Superlat. Microstruct.* **84**, 66 (2015).
23. S. Wang, P. Choubey, Y. X. Chong, W. Chen, W. Ren, H. Eisaki, S. Uchida, P. J. Hirschfeld, and J. C. S. Davis, *Scattering interference signature of a pair density wave state in the cuprate pseudogap phase*, *Nature Commun.* **12**, 6087 (2021).
24. S. F. Kivelson and S. Lederer, *Linking the pseudo-gap in the cuprates with local symmetry breaking: a commentary*, *PNAS* **116**, 14395 (2019).
25. N. J. Robinson, P. D. Johnson, T. M. Rice, and A. M. Tsvelik, *Anomalies in the pseudogap phase of the cuprates: Competing ground states and the role of umklapp scattering*, *Rep. Prog. Phys.* **82**, 126501 (2019).
26. V. Mishra, U. Chatterjee, J. C. Campuzano, and M. R. Norman, *Effect of the pseudogap on the transition temperature in the cuprates and implications for its origin*, *Nat. Phys.* **10**, 357 (2014).

27. R. V. Vovk, Z. F. Nazyrov, G. Ya. Khadzhai, V. M. Pinto Simoes, and V. V. Kruglyak, *Effect of transverse and longitudinal magnetic field on the excess conductivity of* YBa$_2$Cu$_{3-z}$Al$_z$O$_{7-\delta}$ *single crystals with a given topology of plane defects*, *Funct. Mater.* **20**, 208 (2013).

28. B. A. Malik, G. H. Rather, K. Asokan, and M. A. Malik, *Study on excess conductivity in* YBCO + $x$Ag *composites*, *Appl. Phys. A* **127**, 294 (2021).

29. R. I. Rey, C. Carballeira, J. M. Doval, J. Mosqueira, M. V. Ramallo, A. Ramos-Álvarez, D. Sóñora, J. A. Veira, J. C. Verde, and F. Vidal, *The conductivity and the magnetization around $T_c$ in optimally-doped* YBa$_2$Cu$_3$O$_{7-\delta}$ *revisited: quantitative analysis in terms of fluctuating superconducting pairs*, *Supercond. Sci. Technol.* **32**, 045009 (2019).

30. A. L. Solovjov, R. V. Vovk, and K. Rogacki, *Unusual behavior of pseudogap in high-temperature superconductors under the influence of external factors. Part I (Review Article)*, *Low Temp. Phys.* **51**, 1061 (2025) [*Fiz. Nyzk. Temp.* **51**, 1180 (2025)].

31. P. G. De Gennes, *Superconductivity of Metals and Alloys*, W. A. Benjamin, Inc., New York, Amsterdam (1966).

32. E. V. Petrenko, K. Rogacki, A. V. Terekhov, L. V. Bludova, Yu. A. Kolesnichenko, N. V. Shytov, D. M. Sergeyev, E. Lähderanta, and A. L. Solovjov, *Evolution of the pseudogap temperature dependence in* YBa$_2$Cu$_3$O$_{7-\delta}$ *films under the influence of a magnetic field*, *Low Temp. Phys.* **50**, 299 (2024) [*Fiz. Nyzk. Temp.* **50**, 325 (2024)].

33. E. V. Petrenko, L. V. Omelchenko, Yu. A. Kolesnichenko, N. V. Shytov, K. Rogacki, D. M. Sergeyev, and A. L. Solovjov, *Study of fluctuation conductivity in* YBa$_2$Cu$_3$O$_{7-\delta}$ *films in strong magnetic fields*, *Low Temp. Phys.* **47**, 1050 (2021) [*Fiz. Nizk. Temp.* **47**, 1148 (2021)].

34. P. Przyslupski, I. Komissarov, W. Paszkowicz, P. Dluzewski, R. Minikayev, and M. Sawicki, *Structure and magnetic characterization of* La$_{0.67}$Sr$_{0.33}$MnO$_3$/YBa$_2$Cu$_3$O$_7$ *superlattices*, *J. Appl. Phys.* **95**, 2906 (2004).

35. E. V. L. de Mello, M. T. D. Orlando, J. L. Gonzalez, E. S. Caixeiro, and E. Baggio-Saitovich, *Pressure studies on the pseudogap and critical temperatures of a high-$T_c$ superconductor*, *Phys. Rev. B* **66**, 092504 (2002).

36. A. L. Solovjov, H.-U. Habermeier, and T. Haage, *Fluctuation conductivity in* YBa$_2$Cu$_3$O$_{7-y}$ *films of various oxygen content*, II. YBCO *films with* $T_c \gg$ 80 K, *Low Temp. Phys.* **28**, 99 (2002) [*Fiz. Nizk. Temp.* **28**, 144 (2002)].

37. B. Oh, K. Char, A. D. Kent, M. Naito, M. R. Beasley, T. H. Geballe, R. H. Hammond, A. Kapitulnik, and J. M. Graybeal, *Upper critical field, fluctuation conductivity, and dimensionality of* YBa$_2$Cu$_3$O$_7$, *Phys. Rev. B* **37**, 7861 (1988).

38. E. Nazarova, A. Zaleski, and K. Buchkov, *Doping dependence of irreversibility line in* Y$_{1-x}$Ca$_x$Ba$_2$Cu$_3$O$_{7-\delta}$, *Physica C* **470**, 421 (2010).

39. B. P. Stojkovic and D. Pines, *Theory of the longitudinal and Hall conductivities of the cuprate superconductors*, *Phys. Rev. B* **55**, 8576 (1997).

40. F. Rullier-Albenque, H. Alloul, and G. Rikken, *High-field studies of superconducting fluctuations in high-$T_c$ cuprates: Evidence for a small gap distinct from the large pseudogap*, *Phys. Rev. B* **84**, 014522 (2011).

41. W. Lang, G. Heine, P. Schwab, X. Z. Wang, and D. Bauerle, *Paraconductivity and excess Hall effect in epitaxial* YBa$_2$Cu$_3$O$_7$ *films induced by superconducting fluctuations*, *Phys. Rev. B* **49**, 4209 (1994).

42. R. V. Vovk, N. R. Vovk, G. Ya. Khadzhai, I. L. Goulatis, and A. Chroneos, *Effect of praseodymium on the electrical resistance of* YBa$_2$Cu$_3$O$_{7-\delta}$ *single crystals*, *Solid State Commun.* **190**, 18 (2014).

43. R. V. Vovk, N. R. Vovk, G. Ya. Khadzhai, O. V. Dobrovolskiy, and Z. F. Nazyrov, *Effect of high pressure on the fluctuation paraconductivity in* Y$_{0.95}$Pr$_{0.05}$Ba$_2$Cu$_3$O$_{7-\delta}$ *single crystals*, *Curr. Appl. Phys.* **14**, 1779 (2014).

44. R. V. Vovk, G. Ya. Khadzhai, and O. V. Dobrovolskiy, *Resistive measurements of the pseudogap in lightly* Pr*-doped* Y$_{1-x}$Pr$_x$Ba$_2$Cu$_3$O$_{7-\delta}$ *single crystals under high hydrostatic pressure*, *Solid State Commun.* **204**, 64 (2015).

45. R. Peters and J. Bauer, *Local origin of the pseudogap in the attractive Hubbard model*, *Phys. Rev. B* **92**, 014511 (2015).

46. Y. B. Xie, *Superconducting fluctuations in the high-temperature superconductors: Theory of the dc resistivity in the normal state*, *Phys. Rev. B* **46**, 13997 (1992).

47. G. D. Chryssikos, E. I. Kamitsos, J. A. Kapoutsis, A. P. Patsis, V. Psycharis, A. Koufoudakis, Ch. Mitros, G. Kallias, E. Gamari-Seale, and D. Niarchos, *X-ray diffraction and infrared investigation of* RBa$_2$Cu$_3$O$_7$ *and* R$_{0.5}$Pr$_{0.5}$Ba$_2$Cu$_3$O$_7$ *compounds* (R, Y *and lanthanides*), *Physica C* **254**, 44 (1995).

48. S. Hikami and A. I. Larkin, *Magnetoresistance of high-temperature superconductors*, *Mod. Phys. Lett. B* **2**, 693 (1988).

49. A. L. Solovjov, L. V. Omelchenko, R. V. Vovk, O. V. Dobrovolskiy, S. N. Kamchatnaya, and D. M. Sergeev, *Peculiarities in the pseudogap behavior in optimally doped* YBa$_2$Cu$_3$O$_{7-\delta}$ *single crystals under pressure up to 1 GPa*, *Curr. Appl. Phys.* **16**, 931 (2016).

50. L. G. Aslamazov and A. L. Larkin, *The influence of fluctuation pairing of electrons on the conductivity of the normal metal*, *Phys. Lett. A* **26**, 238 (1968).

51. A. L. Solovjov, *Fluctuation conductivity in* Y-Ba-Cu-O *films with artificially produced defects*, *Low Temp. Phys.* **28**, 812 (2002) [*Fiz. Nizk. Temp.* **28**, 1138 (2002)].

52. R. V. Vovk and A. L. Solovjov, *Electric transport and pseudogap in high-temperature superconducting compounds of system 1-2-3 under conditions of all-round compression*, *Low Temp. Phys.* **44**, 81 (2018) [*Fiz. Nizk. Temp.* **44**, 111 (2018)].

53. A. L. Solovjov, L. V. Omelchenko, V. B. Stepanov, R. V. Vovk, H.-U. Habermeier, H. Lochmajer, P. Przysłupski, and K. Rogacki, *Specific temperature dependence of pseudogap in* YBa$_2$Cu$_3$O$_{7-\delta}$ *nanolayers*, *Phys. Rev. B* **94**, 224505 (2016).

54. Y. Matsuda, T. Hirai, S. Komiyama, T. Terashima, Y. Bando, K. Iijima, K. Yamamoto, and K. Hirata, *Magnetoresistance of c-axis-oriented epitaxial* YBa$_2$Cu$_3$O$_{7-x}$ *films above $T_c$*, *Phys. Rev. B* **40**, 5176 (1989).

55. H. Alloul, F. Rullier-Albenque, B. Vignolle, D. Colson, and A. Forget, *Superconducting fluctuations, pseudogap and phase diagram in cuprates*, EPL **91**, 37005 (2010).
56. T. Kondo, A. D. Palczewski, Yoichiro Hamaya, T. Takeuchi, J. S. Wen, Z. J. Xu, G. Gu, and A. Kaminski, *Formation of gapless Fermi arcs and fingerprints of order in the pseudogap state of cuprate superconductors*, Phys. Rev. Lett. **111**, 157003 (2013).
57. S. Badoux, W. Tabis, F. Laliberté, G. Grissonnanche, B. Vignolle, D. Vignolles, J. Béard, D. A. Bonn, W. N. Hardy, R. Liang, N. Doiron-Leyraud, Louis Taillefer, and C. Proust, *Change of carrier density at the pseudogap critical point of a cuprate superconductor*, Nature (London) **531**, 210 (2016).
58. A. A. Kordyuk, *Pseudogap from ARPES experiment: three gaps in cuprates and topological superconductivity*, Low Temp. Phys. **41**, 319 (2015) [*Fiz. Nizk. Temp.* **41**, 417 (2015)].
59. A. L. Solovjov and V. M. Dmitriev, *Resistive studies of the pseudogap in YBCO films with consideration of the transition from BCS to Bose–Einstein condensation*, Low. Temp. Phys. **32**, 99 (2006) [*Fiz. Nizk. Temp.* **32**, 139 (2006)].
60. B. Leridon, A. Defossez, J. Dumont, J. Lesueur, and J. P. Contour, *Conductivity of underdoped* $YBa_2Cu_3O_{7-\delta}$: *evidence for incoherent pair correlations in the pseudogap regime*, Phys. Rev. Lett. **87**, 197007 (2001)
61. V. L. Ginzburg and L. D. Landau, *On the Theory of Superconductivity*, in: *On Superconductivity and Superfluidity*, Springer, Berlin-Heidelberg (2009).
62. E. M. Lifshitz and L. P. Pitaevski, *Statistical Physics*, Nauka (1978).
63. A. Kapitulnik, M. R. Beasley, C. Castellani, and C. Di Castro, *Thermodynamic fluctuations in the high-$T_c$ perovskite superconductors*, Phys. Rev. B **37**, 537 (1988).
64. T. Schneider and J. M. Singer, *Phase Transition Approach to High-Temperature Superconductivity: Universal Properties of Cuprate Superconductors*, Imperial College Press, London (2000).
65. P. Pieri, G. C. Strinati, and D. Moroni, *Magnetic field effect on the pseudogap temperature within precursor superconductivity*, Phys. Rev. Lett. **89**, 127003 (2002).
66. T. Shibauchi, L. Krusin-Elbaum, Ming Li, M. P. Maley, and P. H. Kes, *Closing the pseudogap by Zeeman splitting in* $Bi_2Sr_2CaCu_2O_{8+y}$ *at high magnetic fields*, Phys. Rev. Lett. **86**, 5763 (2001).
67. Y. Yamada, K. Anagawa, T. Shibauchi, T. Fujii, T. Watanabe, A. Matsuda, and M. Suzuki, *Interlayer tunneling spectroscopy and doping-dependent energy-gap structure of the trilayer superconductor* $Bi_2Sr_2Ca_2Cu_3O_{10+\delta}$, Phys. Rev. B **68**, 054533 (2003).
68. J. Stajic, A. Iyengar, K. Levin, B. R. Boyce, and T. R. Lemberger, *Cuprate pseudogap: Competing order parameters or precursor superconductivity*, Phys. Rev. B **68**, 024520 (2003)
69. A. L. Solovjov, L. V. Omelchenko, R. V. Vovk, O. V. Dobrovolskiy, Z. F. Nazyrov, S. N. Kamchatnaya, and D. M. Sergeyev, *Hydrostatic-pressure effects on the pseudogap in slightly doped* $YBa_2Cu_3O_{7-\delta}$ *single crystals*, Physica B **493**, 58 (2016).
70. D. S. Inosov, J. T. Park, A. Charnukha, Y. Li, A. V. Boris, B. Keimer, and V. Hinkov, *Crossover from weak to strong pairing in unconventional superconductors*, Phys. Rev. B **83**, 214520 (2011).
71. Ø. Fischer, M. Kugler, I. Maggio-Aprile, and C. Berthod, *Scanning tunneling spectroscopy of high-temperature superconductors*, Rev. Mod. Phys. **79**, 353 (2007).
72. A. I. D'yachenko, V. Y. Tarenkov, V. V. Kononenko, and E. M. Rudenko, *Effect of pressure on the pseudogap in Bi2223: cuprates are not strongly coupled superconductor*, Metallofiz. Noveishie Tekhnol. **38**, 565 (2016).
73. A. L. Solovjov, E. V. Petrenko, L. V. Omelchenko, R. V. Vovk, I. L. Goulatis, and A. Chroneos, *Effect of annealing on a pseudogap state in untwinned* $YBa_2Cu_3O_{7-\delta}$ *single crystals*, Sci. Rep. **9**, 9274 (2019).
74. Ye Sun and K. Maki, *Impurity effects in d-wave superconductors*, Phys. Rev. B **51**, 6059 (1995).
75. L. M. Ferreira, P. Pureur, H. A. Borges, and P. Lejay, *Effects of pressure on the fluctuation conductivity of* $YBa_2Cu_3O_7$, Phys. Rev. B **69**, 212505 (2004).
76. A. L. Solovjov, L. V. Omelchenko, E. V. Petrenko, R. V. Vovk, V. V. Khotkevych, and A. Chroneos, *Peculiarities of pseudogap in* $Y_{0.95}Pr_{0.05}Ba_2Cu_3O_{7-\delta}$ *single crystals under pressure up to 1.7 GPa*, Sci. Rep. **9**, 20424 (2019).
77. A. L. Solovjov, K. Rogacki, N. V. Shytov, E. V. Petrenko, L. V. Bludova, A. Chroneos, and R. V. Vovk, *Influence of strong electron irradiation on fluctuation conductivity and pseudogap in* $YBa_2Cu_3O_{7-\delta}$ *single crystals*, Phys. Rev. B **111**, 174508 (2025).

---

Вивчення впливу магнітного поля на температурну залежність псевдощілини у оптимально допованих плівках $YBa_2Cu_3O_{7-\delta}$

A. S. Kolisnyk, M. V. Shytov, E. V. Petrenko,
A. V. Terekhov, L. V. Bludova, A. Sedda,
E. Lähderanta, A. L. Solovjov

Проаналізовано вплив магнітного поля $B$ до 9 Тл вздовж осі $c$ ($B \parallel c$) на питомий опір $\rho(T)$, флуктуаційну провідність $\sigma'(T)$ та псевдощілину $\Delta^*(T)$ у тонких плівках $YBa_2Cu_3O_{7-\delta}$ з критичною температурою надпровідного переходу $T_c = 88,8$ К. На відміну від попередньої роботи (Low Temp. Phys. **51**, 1061 (2025) [*Fiz. Nyzk. Temp.* **51**, 1180 (2025)]), де магнітне поле було спрямоване вздовж площини $ab$ ($B \parallel ab$), вплив поля на зразок сильніший завдяки внеску як спін-орбітального, так і зеєманівського ефектів. Як і очікувалося, магнітне поле не впливає на $\rho(T)$ у нормальному стані. Однак, при надпровідному переході магнітне поле $B$ різко збільшує питомий опір $\rho(T)$, ширину надпровідного переходу $\Delta T_c$ та довжину когерентності вздовж осі $c$, $\xi_c(0)$, але водночас зменшує як $T_c$, так і діапазон надпровідних флуктуацій $\Delta T_{\text{fl}}$. Флуктуаційна провідність виявляє перехід при характерній температурі $T_0$: від тривимірної теорії Асламазова–Ларкіна (3D–AL) поблизу $T_c$ до двовимірної теорії флуктуацій Макі–Томпсона (2D–MT). Проте при $B = 1$ Тл внесок 2D–MT повністю пригнічується, і вище $T_0$ залежність $\sigma'(T)$ несподівано описується внеском 2D–AL флуктуацій, що вказує на формування двовимірної вихрової решітки в плівці під дією магнітного поля. Виявлено, що температура переходу BEC–BCS, $T_{pair}$, яка відповідає максимуму залежності $\Delta^*(T)$, зміщується в область нижчих температур зі збільшенням $B$, а максимальне значення $\Delta^*(T_{pair})$ зменшується в полях $B > 5$ Т. Встановлено, що зі збільшенням поля низькотемпературний максимум поблизу $T_0$ розмивається та зникає при $B > 1$ Т. Крім того, вище температури Гінзбурга $T_G$, для $B > 1$ Т, на $\Delta^*(T)$ при $T_{min}$ з'являється мінімум, який стає дуже виразним з подальшим збільшенням $B$. Внаслідок цього загальне значення $\Delta^*(T_G)$ помітно зменшується, найімовірніше, через ефект розриву пар. Водночас $\Delta T_{fl}$ та $\xi_c(0)$ різко зростають приблизно в три рази зі збільшенням $B$ вище 1 Т. Отримані результати підтверджують можливість формування вихрового стану в $YBa_2Cu_3O_{7-\delta}$ магнітним полем та його еволюцію зі збільшенням $B$.

Ключові слова: високотемпературні надпровідники, плівки YBCO, надлишкова провідність, флуктуаційна провідність, магнітне поле, довжина когерентності, псевдощілина.